\documentclass[preprint,11pt]{aastex}
\usepackage{epstopdf}

\shorttitle{A Survey of Merger Remnants II}

\shortauthors{Rothberg \& Joseph}

\begin{document}

\title{A Survey of Merger Remnants II:  The Emerging Kinematic and Photometric Correlations
\footnote{Some of the data
presented herein were obtained at the W.M. Keck Observatory, which is operated as a scientific
partnership among the California Institute of Technology, the University of California and the
National Aeronautics and Space Administration. The Observatory was made possible by the generous
financial support of the W.M. Keck Foundation.}}

\author{B. Rothberg}
\affil{Space Telescope Science Institute, 3700 San Martin Drive Baltimore, MD 21218}
\email{rothberg@stsci.edu}

\author{R. D. Joseph}
\affil{Institute for Astronomy, 2680 Woodlawn Drive,
    Honolulu, HI 96822}

\begin{abstract}
\indent This paper is the second in a series exploring the properties of 51 {\it optically} selected,
single-nuclei merger remnants.  Spectroscopic data have been obtained for a sub-sample of 38 mergers and combined
with previously obtained infrared photometry to test whether mergers exhibit the same correlations as elliptical galaxies
among parameters such as stellar luminosity and distribution, central stellar velocity dispersion ($\sigma$$_{\circ}$), and metallicity.
Paramount to the study is to test whether mergers lie on the Fundamental Plane.  Measurements of $\sigma$$_{\circ}$
have been made using the Ca triplet absorption line at 8500 {\AA} for all 38 mergers in the sub-sample.  Additional
measurements of $\sigma$$_{\circ}$ were made for two of the mergers in the sub-sample using the CO 
absorption line at 2.29 $\micron$.  The results indicate that mergers show a strong correlation among the parameters
of the Fundamental Plane but fail to show a strong correlation between $\sigma$$_{\circ}$ and metallicity (Mg$_{2}$).
In contrast to earlier studies, the $\sigma$$_{\circ}$ of the mergers are consistent with objects which lie somewhere between
intermediate-mass and luminous giant elliptical galaxies.  However, the discrepancies with earlier studies appears to correlate 
with whether the Ca triplet or CO absorption lines are used to derive $\sigma$$_{\circ}$, with the latter almost always producing 
smaller values.   Finally, the photometric and kinematic data are used to demonstrate for the first time that the central 
phase-space density of mergers are equivalent to elliptical galaxies.  This resolves a long-standing criticism of the merger hypothesis.  

\end{abstract}

\section{Introduction}
\indent This paper is the second in a series addressing the viability of the Toomre Merger Hypothesis \markcite{1977egsp.conf..401T}({Toomre} 1977) 
that two merging spiral galaxies can produce a bona fide elliptical galaxy.  The first paper \markcite{2004AJ....128.2098R}({Rothberg} \& {Joseph} 2004), 
hereafter Paper I, used {\it K}-band infrared photometry to determine whether mergers undergo violent relaxation,
as predicted by numerical models, how their structural parameters compare with elliptical galaxies, and investigate
whether mergers have luminosities consistent with  {\it L}$^{*}$ ellipticals.  Fifty-one advanced
merger remnants were {\it optically} selected for the study.  The results showed that stellar light profile of most of the mergers could
be modeled fairly well with a de Vaucouleurs {\it r}$^{1/4}$ profile.  This profile shape is produced as a result of the dissipationless 
collapse of a stellar system during formation \markcite{1982MNRAS.201..939V}({van Albada} 1982).  One way to initiate this is the process of violent relaxation 
\markcite{1967MNRAS.136..101L,1991MNRAS.253..703H}({Lynden-Bell} 1967; {Hjorth} \& {Madsen} 1991), in which the stars are 
scattered by the net gravitational field of the system.  Numerical models by
\markcite{1988ApJ...331..699B,1992ApJ...393..484B}{Barnes} (1988, 1992), among others have shown that mergers undergo violent relaxation, and the net result 
produces a change in the stellar distribution from an exponential light profile, to one in which light approximately follows an {\it r}$^{1/4}$ profile 
(the same profile shape observed in elliptical galaxies).  The results also showed that the mergers are fairly luminous at {\it K}-band,
nearly two-thirds have {\it L} $\geq$ 1 {\it L}$^{*}$$_{ellip}$, and the {\it K}-band isophotal shapes of the mergers appear to
be predominantly disky.\\
\indent  However, photometry alone is insufficient to conclude whether mergers can produce real elliptical galaxies.  While an 
{\it r}$^{1/4}$ profile is consistent with the merger hypothesis, information about the stellar dynamics of the system provides
a stronger test. Observations have shown that spiral bulges, S0 and elliptical galaxies all share a common correlation
between certain kinematic and photometric parameters.  In particular, these objects are known to lie
along a two-dimensional plane, or ``Fundamental Plane,'' embedded in a three-dimensional space comprised of the central velocity
dispersion ($\sigma$$_{\circ}$), half-light or ``effective''-radius ({\it r}$_{\rm eff}$) and the surface brightness
within the effective radius ($<$$\mu$$>$$_{\rm eff}$)  \markcite{1987ApJ...313...59D,
1992ApJ...399..462B}({Djorgovski} \& {Davis} 1987; {Bender}, {Burstein}, \&  {Faber} 1992).  While the small scatter in the Fundamental Plane makes it useful as a distance indicator,
the tight correlation among the observed parameters provides important clues about the underlying
physics of these systems.  A similar relationship to the Fundamental Plane can be derived using the Virial Theorem,
assuming that {\it M/L} $\propto$ {\it L}$^{\alpha}$ is constant, and that elliptical galaxies and bulges form a homologous family
in their scaling relations.  This suggests that the empirically derived Fundamental Plane is a representation of the Virial Theorem.
If mergers lie on or near the Fundamental Plane, this suggests that they are physically similar to elliptical galaxies.\\
\indent Elliptical galaxies also show a strong correlation between $\sigma$$_{\circ}$ and Mg$_{2}$, which serve as a proxy for
mass and metallicity \markcite{1981ApJ...246..680T,1981MNRAS.196..381T}({Tonry} \& {Davis} 1981; {Terlevich} {et~al.} 1981).  The deeper the potential well, the more the stellar
populations are metal enriched.  The kinematic data presented here allow this relationship to be tested for the first time
among mergers.  It has been shown that mergers undergo strong star-formation, \markcite{1987nngp.proc...18S,1985MNRAS.214...87J}({Schweizer} 1987; {Joseph} \& {Wright} 1985) likely 
triggered by the dissipation of gas
from the progenitor galaxies.  A test of the mass-metallicity relation may provide constraints on the stellar populations
present, or forming in mergers and how they compare with those in elliptical galaxies.\\
 \indent The kinematic data presents a unique opportunity to directly address an early, but
serious criticism of the mergers-make-ellipticals hypothesis;  that the stellar phase-space densities of 
elliptical and spiral galaxies are very different.  \markcite{1986ApJ...310..593C}{Carlberg} (1986) used surface photometry of 
galaxy cores in conjunction with velocity dispersions to compare the phase-space densities of elliptical and spiral galaxies.  The results showed 
that all but the brightest ellipticals have significantly higher phase-space densities
than spirals.  According to the collisionless Boltzmann equation, the phase-space density of a stellar system 
remains constant or decreases.  It cannot increase.  Therefore, most elliptical galaxies cannot form from 
purely dissipationless (stellar-only) mergers.  This led \markcite{1987nngp.proc.....F}{Gunn} (1987) to comment,
 ``I do not think you can make rocks by merging clouds.''  \markcite{1993ApJ...416..415H}{Hernquist}, {Spergel}, \&  {Heyl} (1993) 
explored the evolution of phase-space densities in mergers using numerical simulations.  The results were consistent with Carlberg's
findings.  Stellar disks lack sufficient material at high-phase space densities to form elliptical 
galaxies.  One way to overcome this is via a starburst 
induced by dissipative collapse of the gas.  \markcite{1991ApJ...370L..65B}{Barnes} \& {Hernquist} (1991), among others, used hydrodynamical
simulations to show that tidal forces in a merger can drive a significant fraction of the gas into the
central region.  The gas is not subject to the collisionless Boltzmann equation and becomes more centrally
concentrated than the stars.  The infall can fuel a starburst, producing more stars and increasing the
phase-space density of the merger.  \markcite{1994ApJ...437L..47M}{Mihos} \& {Hernquist} (1994) included the effects of dissipative collapse
in their numerical simulations of mergers.  Their models predict that when the gas falls to the center of the
merger, a strong starburst is triggered.  The resulting central starburst should affect the
shape of the surface brightness profile, producing a sharp increase in luminosity near the nucleus.  The results from Paper I
showed that nearly one-third of the mergers show evidence of just such an increase in luminosity near the nucleus,
producing surface brightness profiles that rise significantly above an {\it r}$^{1/4}$ profile.  The kinematic data presented here
make it possible for the first time to directly measure the phase-space density of mergers and compare them with spiral and elliptical galaxies.\\
\indent To date, few kinematic studies of mergers have been undertaken.  The largest, and most in-depth kinematic study 
was conducted by \markcite{1986ApJ...310..605L}{Lake} \& {Dressler} (1986). They measured the $\sigma$$_{\circ}$ of 13 mergers 
from the Atlas of Peculiar Galaxies \markcite{1966apga.book.....A}({Arp} 1966) and A Catalogue of Southern Peculiar Galaxies 
\markcite{1987cspg.book.....A}({Arp} \& {Madore} 1987) using both the \ion{Mg}{1}{\it b} ($\lambda$ $\sim$ 5170 {\AA}) and Ca triplet 
($\lambda$ $\sim$ 8500 {\AA}) stellar absorption lines.  The central velocity dispersions were then plotted against {\it M$_{B}$} to test 
whether they fit the Faber-Jackson relation \markcite{1976ApJ...204..668F}({Faber} \& {Jackson} 1976).  This relation is a correlation between the 
total luminosity, {\it L}, and $\sigma$$_{\circ}$, such that {\it L} $\propto$ $\sigma$$^{4}$.  It is similar to the Fundamental Plane,
but the relation produces a larger scatter, implying the presence of second parameter.
Lake \& Dressler predicted that the mergers should be 1.5 mag brighter, with 
$\sigma$$_{\circ}$ $\sim$ 30$\%$ smaller than elliptical galaxies.  The results showed that the mergers
were only $\sim$ 0.4 mag brighter in luminosity for a given $\sigma$$_{\circ}$.
The differences between the mergers and ellipticals in the Coma and Virgo clusters were not outside 
the range of differences seen between other samples of elliptical galaxies.  They also found that the 
\ion{Mg}{1}{\it b} absorption lines were more problematic than the Ca triplet lines for deriving 
$\sigma$$_{\circ}$.  Contamination from nearby emission lines, most likely caused by star-formation 
and/or young stars, can interfere with the \ion{Mg}{1}{\it b} lines.  Lake \& Dressler suggested that 
using the {\it H}-band to measure the luminosity and the Ca triplet lines to derive $\sigma$$_{\circ}$ 
would be more effective because they are less likely to be affected by star-formation, and are more 
sensitive to the older stellar population.\\
\indent   No further studies of mergers using the Ca triplet absorption line to derive velocity dispersions 
have been made, and no studies have ever been conducted using the Ca triplet line and the Fundamental Plane.
Later studies have substituted the use of the Ca triplet line with the CO absorption line at 2.29 $\micron$ (and sometimes at 
1.63 $\micron$) to study whether mergers lie on 
or near the Fundamental Plane (see \markcite{1995PASP..107...68G}{Gaffney}, {Lester}, \&  {Doppmann} (1995) for a discussion of the use of the
2.29 $\micron$ CO absorption line for kinematic studies).  The switch to the CO
line was motivated by studies of Luminous and Ultra-Luminous Infrared Galaxies.  LIRGs and ULIRGs are objects with {\it L$_{FIR}$} $\geq$ 
10$^{11}${\it L$_{\sun}$} and 10$^{12}${\it L$_{\sun}$} respectively, where {\it L$_{FIR}$} is measured from 8-1000 $\micron$ (see \markcite{1996ARA&A..34..749S}{Sanders} \& {Mirabel} (1996)
for a more in-depth look).  A vast majority of these objects have disturbed morphologies consistent with an ongoing merger.   
The CO line was chosen primarily to avoid the effects of high optical extinction caused by dust in LIRG/ULIRGs.\\
\indent Several LIRG/ULIRG merger studies been made using the 2.29 $\micron$ CO absorption line to
obtain velocity dispersions, including \markcite{1994ApJ...437L..23D}{Doyon} {et~al.} (1994), 
\markcite{1998ApJ...497..163S}{Shier} \& {Fischer} (1998), and \markcite{1999MNRAS.309..585J}{James} {et~al.} (1999). \markcite{1995A&A...301...55O}{Oliva} {et~al.} (1995)
and \markcite{2001ApJ...563..527G}{Genzel} {et~al.} (2001) used both the 2.29 $\micron$ and 1.63 $\micron$ CO absorption line.  However, a 
major difference between these studies and the one presented here lies in the sample selection.  
The mergers presented in this paper have been selected based on {\it optical} morphology alone, and 
not based on their infrared luminosity.  In addition, all of the mergers in this paper have only one nucleus, 
in contrast to earlier LIRG/ULIRG kinematic studies, which included objects with two nuclei.  
Of the 38 mergers in this sample for which velocity dispersions have been obtained, only 8 have been previously measured using the Ca triplet,
and 8 have been previously measured using the infrared CO absorption feature.  One object,
NGC 7252, has been studied before using both the Ca triplet and CO studies, 
mainly because it is considered a ``canonical'' merger.\\
\indent  {\it K}-band photometry was selected for the survey for two reasons.  First, dust extinction is $\sim$ 1/10$^{th}$
that at {\it V}-band.  Second, the blackbody emission of older, late-type stars, which trace the majority of the stellar mass in
both elliptical and spiral galaxies, peaks in the near-infrared beyond 1 $\micron$ \markcite{1981IAUS...96..297A}({Aaronson} 1981). \\
\indent However, one problem with using the {\it K}-band is the possibility of contamination from
Asymptotic Giant Branch (AGB) stars.  These stars also peak in luminosity in the near-infrared.
\markcite{2002A&A...393..149M}{Mouhcine} \& {Lan{\c c}on} (2002) modeled the contribution of AGB stars to the total {\it K}-band luminosity for a single-burst population.  
They found that the contribution evolves rapidly from a few percent at $\simeq$ 0.1 Gyr to $\simeq$ 60$\%$ of the light at 0.6-0.7 Gyr
and then quickly declines soon after to $\leq$ 30$\%$.  The contribution to the integrated light is also directly proportional to metallicity.  
Yet, the contribution from AGB stars to the total {\it K}-band light is still less than the contribution of young stars to the integrated light
at optical wavelengths.\\
\indent The Ca triplet lines at 8498.0, 8542.1, and 8662.1 {\AA} were selected as the primary tool to measure the stellar velocity dispersions
because they are strong, well-separated, and narrow \markcite{1984ApJ...286...97D}({Dressler} 1984).  These lines are particularly strong in late-type giants.
Other optical lines are often broader, too complex, or very sensitive to template mismatch.  Absorption lines at shorter wavelengths 
are also more likely to be contaminated by strong nearby emission lines from in galaxies undergoing star-formation or
with young stellar populations.  Finally, the more red-ward Ca triplet lines are somewhat less affected by extinction than 
absorption lines at shorter wavelengths.\\
\indent Velocity dispersions have also been measured using the CO absorption feature at 2.29 $\micron$ for
two galaxies in the sample, NGC 1614 and NGC 2623.  These mergers are both classified as LIRGs.
As noted earlier, the selection of the CO stellar absorption line is primarily motivated by the desire to avoid as much extinction
as possible because LIRG/ULIRGs are known to be quite dusty.  The CO line at 2.29 $\micron$ is quite strong and well-defined.
However, there is one serious drawback. The CO absorption line is sensitive not only to late-type giant stars, but the presence 
of red super-giants and strong starbursts.  Velocity dispersions derived from this line may not be probing the same late-type
stars as other stellar absorption lines.  Furthermore, the presence of hot-dust from a starburst, or the presence of an AGN
can ``fill-in'' the CO line and affect the derived velocity dispersions.  There is already some evidence to suggest that velocity dispersions
measured from the CO absorption line do not match those measured from lines at shorter wavelengths \markcite{2003AJ....125.2809S}{Silge} \& {Gebhardt} (2003) found that, 
for a sample of 25 early-type galaxies, velocity dispersions obtained from CO were generally smaller, 
sometimes by as much as 40$\%$.  A similar result was found for the mergers and will be discussed subsequently.\\
\indent An important part of comparing the photometric and kinematic properties of mergers with those
of elliptical galaxies rests on assuring that the comparison is made using the same type of diagnostic tools.
Until recently, a comprehensive infrared study of elliptical galaxies did not exist.  The first large survey 
was done by \markcite{1999MNRAS.304..225M}{Mobasher} {et~al.} (1999) in which they constructed a {\it K}-band fundamental plane 
for 48 early-type galaxies in Coma.  \markcite{1999ApJS..124..127P}{Pahre} (1999) 
(hereafter P99), conducted a {\it K}-band photometric study of 251 early-type galaxies from several clusters, 
groups, and in the field.  Using this data along with velocity
dispersions and Mg$_{2}$ measurements obtained from the literature, \markcite{1998AJ....116.1591P}{Pahre}, {Djorgovski}, \& {de  Carvalho} (1998) 
(hereafter P98) conducted an in-depth analysis of several elliptical galaxy correlations,
including the Fundamental Plane.  The {\it K}-band was selected
for these studies primarily for the same reasons as the merger study presented in this paper.  An additional
reason discussed by P98 is that the light at {\it K}-band is less sensitive to metallicity effects.
If the slope of the Fundamental Plane is due to variations in the metallicity of the stellar populations, 
then it should be very different at {\it K}-band compared to optical wavelengths.  If the slope of 
the Fundamental Plane is influenced by structural, or kinematic non-homology
in ellipticals, then the slope at {\it K}-band should be similar to the slope at optical wavelengths.  Both \markcite{1999MNRAS.304..225M}{Mobasher} {et~al.} (1999)
and P98 found the slope of the {\it K}-band Fundamental Plane to be similar to its optical counterparts.\\
\indent The P99 data and P98 results will be used as the basis for the comparison between elliptical galaxies
and the mergers presented here.  Every effort has been made to measure the mergers using the same
techniques and aperture sizes used in the P99 and P98 papers.  All metric measurements in the paper
assume a value of {\it H}$_{\circ}$ = 75 km s$^{-1}$ Mpc$^{-1}$.  An additional note about references in 
this paper; P99 refers to the elliptical galaxy data presented in \markcite{1999ApJS..124..127P}{Pahre} (1999) such as 
log {\it R}$_{\rm eff}$, $\sigma$$_{\circ}$, $<$$\mu$$_{K}$$>$$_{\rm eff}$, Mg$_{2}$, or values derived from the data, like the 
effective phase-space density (f$_{\rm eff}$), while P98 refers to the analysis of the ellipticals
in regards to the fitting of parameter correlations such as the Fundamental Plane
as presented in \markcite{1998AJ....116.1591P}{Pahre} {et~al.} (1998).

\section{Sample}
\subsection{{\it K}-band imaging}
\indent The 51 objects in the sample presented here were optically selected primarily from the Atlas of 
Peculiar Galaxies \markcite{1966apga.book.....A}({Arp} 1966), A Catalogue of Southern Peculiar Galaxies 
\markcite{1987cspg.book.....A}({Arp} \& {Madore} 1987), the Atlas and Catalog of Interacting Galaxies \markcite{1959VV....C......0V}({Vorontsov-Velyaminov} 1959),
and the Uppsala General Catalogue \markcite{1973ugcg.book.....N}({Nilson} 1973).  The specific selection process
is detailed in Paper I.
The selections were based strictly on optical morphology and according to the following criteria:\\
\indent 1)  Tidal tails, loops, and shells, which are induced by strong gravitational
            interaction.\\
\indent 2)  A single nucleus, which, based on numerical studies,
        marks the completion of the merger.  This criteria is important because it marks the point
        at which the merger should begin to exhibit properties in common with elliptical galaxies.\\
\indent 3)  The absence of nearby companions which may induce the presence of tidal
        tails and make the object appear to be in a more advanced stage of merging.\\
\indent 4)  The mergers must be observable from Mauna Kea.  This limited the survey to objects
        with declinations $\geq$ -50$\degr$.\\
\indent The sample is listed in Table 1 and includes names, right ascension, declination and whether there
is photometric and/or spectroscopic data for that object.
Since most of the objects have multiple designations, all subsequent references
to sample galaxies within the paper, tables and figures will first use the NGC designation if available,  
followed by the Arp or Arp-Madore (AM), UGC, VV and lastly the IC designation if no other designation 
is available.  Unless otherwise noted, the galaxies are listed in order of Right Ascension in 
tables and figures.  One object from this sample, AM 0612-373, has no previously recorded redshift.
The redshift used to determine a distance to this galaxy was taken from the Ca triplet measurements
presented in this paper.\\
\clearpage
{
\begin{deluxetable}{llcccc}
\tabletypesize{\normalsize}
\setlength{\tabcolsep}{0.1in}
\tablewidth{0pt}
\tablenum{1}
\pagestyle{empty}
\tablecaption{Merger Sample}
\tablecolumns{6}
\tablehead{
\colhead{Merger Names} &
\colhead{Other Names} &
\colhead{R.A.} &
\colhead{Dec.} &
\colhead{notes} &
\colhead{data\tablenotemark{a}}\\
\colhead{} &
\colhead{} &
\colhead{(J2000)}&
\colhead{(J2000)}&
\colhead{} &
\colhead{}
}
\startdata
UGC 6       &VV 806                     &00$^{h}$ 03$^{m}$ 09$^{s}$ &21$^{\circ}$ 57$^{'}$ 37$^{''}$  &LIRG   &I,S\\
NGC 34      &VV 850                     &00$^{h}$ 11$^{m}$ 06$^{s}$ &-12$^{\circ}$ 06$^{'}$ 26$^{''}$ &LIRG   &I,S\\
Arp 230     &IC 51                      &00$^{h}$ 46$^{m}$ 24$^{s}$ &-13$^{\circ}$ 26$^{'}$ 32$^{''}$ &Shell  &I\\
NGC 455     &Arp 164, UGC 815           &01$^{h}$ 15$^{m}$ 57$^{s}$ &05$^{\circ}$ 10$^{'}$ 43$^{''}$  &       &I,S\\
NGC 828     &UGC 1655                   &02$^{h}$ 10$^{m}$ 09$^{s}$ &39$^{\circ}$ 11$^{'}$ 25$^{''}$  &LIRG   &I\\
UGC 2238    &\nodata                    &02$^{h}$ 46$^{m}$ 17$^{s}$ &13$^{\circ}$ 05$^{'}$ 44$^{''}$  &LIRG   &I\\
NGC 1210    &AM 0304-255                &03$^{h}$ 06$^{m}$ 45$^{s}$ &-25$^{\circ}$ 42$^{'}$ 59$^{''}$ &Shell  &I,S\\
AM 0318-230 &\nodata                    &03$^{h}$ 20$^{m}$ 40$^{s}$ &-22$^{\circ}$ 55$^{'}$ 53$^{''}$ &       &I\\
NGC 1614    &Arp 186                    &04$^{h}$ 33$^{m}$ 59$^{s}$ &-08$^{\circ}$ 34$^{'}$ 44$^{''}$ &LIRG   &I,S\\
Arp 187     &\nodata                    &05$^{h}$ 04$^{m}$ 53$^{s}$ &-10$^{\circ}$ 14$^{'}$ 51$^{''}$ &       &I\\
AM 0612-373 &\nodata                    &06$^{h}$ 13$^{m}$ 47$^{s}$ &-37$^{\circ}$ 40$^{'}$ 37$^{''}$ &       &I,S\\
NGC 2418    &Arp 165, UGC 3931          &07$^{h}$ 36$^{m}$ 37$^{s}$ &17$^{\circ}$ 53$^{'}$ 02$^{''}$  &       &I,S\\
UGC 4079    &\nodata                    &07$^{h}$ 55$^{m}$ 06$^{s}$ &55$^{\circ}$ 42$^{'}$ 13$^{''}$  &       &I\\
NGC 2623    &Arp 243, UGC 4509, VV 79   &08$^{h}$ 38$^{m}$ 24$^{s}$ &25$^{\circ}$ 45$^{'}$ 17$^{''}$  &LIRG   &I,S\\
UGC 4635    &\nodata                    &08$^{h}$ 51$^{m}$ 54$^{s}$ &40$^{\circ}$ 50$^{'}$ 09$^{''}$  &       &I,S\\
NGC 2655    &Arp 225, UGC 4637          &08$^{h}$ 55$^{m}$ 37$^{s}$ &78$^{\circ}$ 13$^{'}$ 23$^{''}$  &Shell  &I,S\\
NGC 2744    &UGC 4757, VV 612           &09$^{h}$ 04$^{m}$ 38$^{s}$ &18$^{\circ}$ 27$^{'}$ 37$^{''}$  &       &I\\
NGC 2782    &Arp 215, UGC 4862          &09$^{h}$ 14$^{m}$ 05$^{s}$ &40$^{\circ}$ 06$^{'}$ 49$^{''}$  &LIRG   &I,S\\
NGC 2914    &Arp 137, UGC 5096          &09$^{h}$ 34$^{m}$ 02$^{s}$ &10$^{\circ}$ 06$^{'}$ 31$^{''}$  &       &I,S\\
UGC 5101    &\nodata                    &09$^{h}$ 35$^{m}$ 51$^{s}$ &61$^{\circ}$ 21$^{'}$ 11$^{''}$  &ULIRG  &I,S\\
AM 0956-282 &\nodata                    &09$^{h}$ 58$^{m}$ 46$^{s}$ &-28$^{\circ}$ 37$^{'}$ 19$^{''}$ &       &I\\
NGC 3256    &AM 1025-433, VV 65         &10$^{h}$ 27$^{m}$ 51$^{s}$ &-43$^{\circ}$ 54$^{'}$ 14$^{''}$ &LIRG   &I,S\\
NGC 3310    &Arp 217, UGC 5786, VV 356  &10$^{h}$ 38$^{m}$ 45$^{s}$ &53$^{\circ}$ 30$^{'}$ 05$^{''}$  &       &I\\
Arp 156     &UGC 5814                   &10$^{h}$ 42$^{m}$ 38$^{s}$ &77$^{\circ}$ 29$^{'}$ 41$^{''}$  &       &I,S\\
NGC 3597    &AM 1112-232                &11$^{h}$ 14$^{m}$ 41$^{s}$ &-23$^{\circ}$ 43$^{'}$ 39$^{''}$ &       &I,S\\
NGC 3656    &Arp 155, UGC 6403, VV 22a  &11$^{h}$ 23$^{m}$ 38$^{s}$ &53$^{\circ}$ 50$^{'}$ 30$^{''}$  &Shell  &I,S\\
NGC 3921    &Arp 224, UGC 6823, VV 31   &11$^{h}$ 51$^{m}$ 06$^{s}$ &55$^{\circ}$ 04$^{'}$ 43$^{''}$  &       &I,S\\
NGC 4004    &UGC 6950, VV 230           &11$^{h}$ 58$^{m}$ 05$^{s}$ &27$^{\circ}$ 52$^{'}$ 44$^{''}$  &       &I,S\\
AM 1158-333 &\nodata                    &12$^{h}$ 01$^{m}$ 20$^{s}$ &-33$^{\circ}$ 52$^{'}$ 36$^{''}$ &       &I\\
NGC 4194    &Arp 160, UGC 7241, VV 261  &12$^{h}$ 14$^{m}$ 09$^{s}$ &54$^{\circ}$ 31$^{'}$ 36$^{''}$  &LIRG   &I,S\\
NGC 4441    &UGC 7572                   &12$^{h}$ 27$^{m}$ 20$^{s}$ &64$^{\circ}$ 48$^{'}$ 06$^{''}$  &       &I,S\\
UGC 8058    &Mrk 231                    &12$^{h}$ 56$^{m}$ 14$^{s}$ &56$^{\circ}$ 52$^{'}$ 25$^{''}$  &ULIRG  &I\\
AM 1255-430 &\nodata                    &12$^{h}$ 58$^{m}$ 08$^{s}$ &-43$^{\circ}$ 19$^{'}$ 47$^{''}$ &       &I,S\\
AM 1300-233 &\nodata                    &13$^{h}$ 02$^{m}$ 52$^{s}$ &-23$^{\circ}$ 55$^{'}$ 18$^{''}$ &LIRG   &I\\
NGC 5018    &UGCA 335                   &13$^{h}$ 13$^{m}$ 00$^{s}$ &-19$^{\circ}$ 31$^{'}$ 05$^{''}$ &Shell  &I,S\\
Arp 193     &UGC 8387, VV 821, IC 883   &13$^{h}$ 20$^{m}$ 35$^{s}$ &34$^{\circ}$ 08$^{'}$ 22$^{''}$  &LIRG   &I,S\\
AM 1419-263 &\nodata                    &14$^{h}$ 22$^{m}$ 06$^{s}$ &-26$^{\circ}$ 51$^{'}$ 27$^{''}$ &       &I,S\\
UGC 9829    &VV 847                     &15$^{h}$ 23$^{m}$ 01$^{s}$ &-01$^{\circ}$ 20$^{'}$ 50$^{''}$ &       &I,S\\
NGC 6052    &Arp 209, UGC 10182, VV 86  &16$^{h}$ 05$^{m}$ 12$^{s}$ &20$^{\circ}$ 32$^{'}$ 32$^{''}$  &       &I,S\\
UGC 10607   &VV 852, IC 4630            &16$^{h}$ 55$^{m}$ 09$^{s}$ &26$^{\circ}$ 39$^{'}$ 46$^{''}$  &       &I,S\\
UGC 10675   &VV 805                     &17$^{h}$ 03$^{m}$ 15$^{s}$ &31$^{\circ}$ 27$^{'}$ 29$^{''}$  &       &I,S\\
NGC 6598    &UGC 11139                  &18$^{h}$ 08$^{m}$ 56$^{s}$ &69$^{\circ}$ 04$^{'}$ 04$^{''}$  &       &I\\
AM 2038-382 &\nodata                    &20$^{h}$ 41$^{m}$ 13$^{s}$ &-38$^{\circ}$ 11$^{'}$ 36$^{''}$ &       &I,S\\
AM 2055-425 &\nodata                    &20$^{h}$ 58$^{m}$ 26$^{s}$ &-42$^{\circ}$ 39$^{'}$ 00$^{''}$ &LIRG   &I,S\\
NGC 7135    &AM 2146-350, IC 5136       &21$^{h}$ 49$^{m}$ 46$^{s}$ &-34$^{\circ}$ 52$^{'}$ 35$^{''}$ &       &I,S\\
UGC 11905   &\nodata                    &22$^{h}$ 05$^{m}$ 54$^{s}$ &20$^{\circ}$ 38$^{'}$ 22$^{''}$  &       &I,S\\
NGC 7252    &Arp 226, AM 2217-245       &22$^{h}$ 20$^{m}$ 44$^{s}$ &-24$^{\circ}$ 40$^{'}$ 41$^{''}$ &       &I,S\\
AM 2246-490 &\nodata                    &22$^{h}$ 49$^{m}$ 39$^{s}$ &-48$^{\circ}$ 50$^{'}$ 58$^{''}$ &ULIRG  &I,S\\
IC 5298     &\nodata                    &23$^{h}$ 16$^{m}$ 00$^{s}$ &25$^{\circ}$ 33$^{'}$ 24$^{''}$  &LIRG   &I,S\\
NGC 7585    &Arp 223                    &23$^{h}$ 18$^{m}$ 01$^{s}$ &-04$^{\circ}$ 39$^{'}$ 01$^{''}$ &Shell  &I,S\\
NGC 7727    &Arp 222, VV 67             &23$^{h}$ 39$^{m}$ 53$^{s}$ &-12$^{\circ}$ 17$^{'}$ 35$^{''}$ &       &I,S\\
\cutinhead{Elliptical}
NGC 5812    &UGCA 398                   &15$^{h}$ 00$^{m}$ 55$^{s}$ &-07$^{\circ}$ 27$^{'}$ 26$^{''}$ &E0     &I,S\\
\enddata
\tablecomments{(a) I = {\it K}-band imaging, S = Spectra}
\end{deluxetable}

}
\clearpage
\indent In addition to the mergers in the sample, an E0 elliptical galaxy, NGC 5812, was also observed.
This object was selected as a ``control-sample '' elliptical.  It was observed using the same
instrument setups and analyzed in exactly the same way as the mergers presented in this paper.
P99 also observed NGC 5812 at {\it K}-band, and extracted $\sigma$$_{\circ}$ and Mg$_{2}$ values
from the literature.  This makes it useful for comparing the P99 photometric results 
with those presented in this paper and ruling out any systematic differences which may affect comparisons
between that study and this one.  The photometric differences between P99 and
the photometry presented here are small; $\Delta$$<$$\mu$$_{K}$$>$$_{\rm eff}$ = 0.01,
$\Delta$ log {\it R}$_{\rm eff}$ = 0.03, $\Delta${\it M}$_{K}$ = 0.02.\\
\indent Within the merger sample there are three sub-samples, ``shell ellipticals,'' 
ultra-luminous and luminous infrared galaxies (LIRG/ULIRGs), and ``normal'' mergers, which are defined 
simply as those galaxies which are neither LIRG/ULIRGs nor shell ellipticals.  Each of the sub-samples
are plotted with different symbols in all figures.  A more in-depth discussion of
the first two sub-samples is given in Paper I.

\subsection{Spectroscopy}
\indent The optical spectroscopic observations consist of a sub-sample of 38 objects taken from the 
{\it K}-band imaging sample.  The sub-sample was not selected based on any particular criteria.  
The goal of the spectroscopic campaign 
was to obtain velocity dispersions for every galaxy in the sample.  Unfortunately, due to 
limitations of time, only 38/51 objects in the sample could be observed.  The 38 objects 
were observed in no particular order.  In addition to the optical spectra, two mergers,
NGC 1614 and NGC 2623, were observed in the infrared at 2.29 $\micron$ 
which corresponds to the CO molecular absorption line.  

\section{Observations}
\subsection{{\it K}-band imaging}
\indent Near-infrared images were obtained using the Quick Infrared Camera (QUIRC) 
1024 $\times$ 1024 pixel HgCdTe infrared array \markcite{1996NewA....1..177H}({Hodapp} {et~al.} 1996) at f/10 
focus on the University of Hawaii 2.2 meter telescope. 
The field of view of the QUIRC array is 193$\arcsec$ x 193$\arcsec$  with a plate scale 
of 0.189$\arcsec$ pixel$^{-1}$.  Exposures from QUIRC were obtained 
by nodding between the source and a blank field of sky.  The on-target observations
were dithered so as to remove any array defects in the final processing.
The infrared photometric standards were selected based on proximity to the 
targets in Right Ascension and Declination. Multiple observations of standard stars were taken 
before and after target observations.  Standards were selected from \markcite{1998AJ....116.2475P}{Persson} {et~al.} (1998), 
\markcite{1995MNRAS.276..734C}{Carter} \& {Meadows} (1995), \markcite{1998AJ....115.2594H}{Hunt} {et~al.} (1998), and \markcite{2001MNRAS.325..563H}{Hawarden} {et~al.} (2001), 
which includes the UKIRT faint standards.  The median seeing for the observations was 0$\arcsec$.8\\
\indent  The {\it K} filter used in this survey for all but two objects conforms to the new 
Mauna Kea Infrared Filter Set \markcite{2002PASP..114..180T}({Tokunaga}, {Simons}, \&  {Vacca} 2002).  One object, UGC 6 was 
observed with the {\it K'} filter \markcite{1992AJ....103..332W}({Wainscoat} \& {Cowie} 1992) and has been converted to {\it K} 
using their conversion equation. Another object, IC 5298, was observed with an older {\it K} 
filter with similar properties to the Mauna Kea {\it K} filter.  No conversion was made 
and it is assumed that {\it K$_{old}$} $\simeq$ {\it K$_{MaunaKea}$}.  Table 2 lists the observation
log for each object, including the total integration time and the measured seeing for each object.\\
\clearpage
{
\begin{deluxetable}{lcc}
\tabletypesize{\normalsize}
\setlength{\tabcolsep}{0.1in}
\tablewidth{0pt}
\tablenum{2}
\pagestyle{empty}
\tablecaption{{\it K}-band Photometry Observation Log}
\tablecolumns{3}
\tablehead{
\colhead{Merger Name} &
\colhead{Integration Time} &
\colhead{seeing}\\
\colhead{} &
\colhead{(sec)}&
\colhead{(\arcsec)}
}
\startdata
UGC 6       &3210 &0.90 \\
NGC 34      &3300 &0.94 \\
Arp 230     &2325 &1.45 \\
NGC 455     &2250 &0.90 \\
NGC 828     &2580 &1.19 \\
UGC 2238    &3480 &0.98 \\
NGC 1210    &2820 &0.68 \\
AM 0318-230 &3660 &0.98 \\
NGC 1614    &2730 &0.64 \\
Arp 187     &3150 &0.92 \\
AM 0612-373 &2160 &1.09 \\
NGC 2418    &3600 &0.68 \\
UGC 4079    &3600 &0.58 \\
NGC 2623    &3390 &0.64 \\
UGC 4635    &3480 &0.64 \\
NGC 2655    &1800 &0.92 \\
NGC 2744    &3600 &0.60 \\
NGC 2782    &3000 &0.62 \\
NGC 2914    &1800 &0.81 \\
UGC 5101    &2700 &1.02 \\
AM 0956-282 &3600 &0.62 \\
NGC 3256    &2925 &1.02 \\
NGC 3310    &2940 &0.90 \\
Arp 156     &2400 &0.86 \\
NGC 3597    &3300 &0.81 \\
NGC 3656    &3150 &0.58 \\
NGC 3921    &3645 &0.88 \\
NGC 4004    &1560 &0.86 \\
AM 1158-333 &3600 &0.69 \\
NGC 4194    &3600 &0.62 \\
NGC 4441    &1440 &0.86 \\
UGC 8058    &1170 &0.62 \\
AM 1255-430 &3420 &0.90 \\
AM 1300-233 &1200 &0.62 \\
NGC 5018    &2025 &0.66 \\
Arp 193     &2625 &0.62 \\
AM 1419-263 &3600 &0.79 \\
UGC 9829    &3600 &0.62 \\
NGC 6052    &3600 &0.71 \\
UGC 10607   &3360 &0.75 \\
UGC 10675   &3600 &0.62 \\
NGC 6598    &2700 &0.66 \\
AM 2038-382 &1920 &0.69 \\
AM 2055-425 &2880 &1.02 \\
NGC 7135    &2520 &0.69 \\
UGC 11905   &3000 &0.81 \\
NGC 7252    &3360 &0.90 \\
AM 2246-490 &2520 &1.17 \\
IC 5298     &3240 &0.62 \\
NGC 7585    &2040 &0.92 \\
NGC 7727    &3240 &0.83 \\
\cutinhead{Elliptical}
NGC 5812    &2400 &1.41 \\

\enddata
\end{deluxetable}

}
\clearpage
\subsection{Optical Spectroscopy}
\indent The optical spectroscopic observations were obtained with the Echellette Spectrograph and 
Imager \markcite{2002PASP..114..851S}({Sheinis} {et~al.} 2002) at the W. M. Keck-II 10-meter observatory. 
The spectrograph covers the wavelength range from 3927-11068 {\AA}.  The observations were 
made in the echelle mode using the 0{$\arcsec$}.5 $\times$ 20{$\arcsec$} slit.  In this mode, 
ten orders are projected onto a 2048$\times$4096 CCD.  ESI has a resolution of 
11.2 km s$^{-1}$ pixel$^{-1}$.  The 0{$\arcsec$}.5 slit projects to 3.24 pixels, providing 
a resolution of {\it R} $\simeq$ 36.2 km s$^{-1}$ or {\it R} $\simeq$ 8200.  
The slit Position Angle was placed along the major axis of each galaxy. The P.A. of the major axis
was measured using the {\it K}-band data and is listed for each galaxy in the spectroscopic observation
log of Table 3, along with the total integration time for each object.  Dome flats and calibration 
arcs were observed at the beginning and end of each night.  Dome flats were selected over internal 
flats for use in the reduction process in order to correct the high-frequency fringing which 
occurs in the higher orders of the spectra (beyond $\lambda$ $\simeq$ 7100 {\AA}). The calibration 
arcs consisted of separate exposures containing Cu-Ar lines and Hg-Xe-Ne lines.  Spectrophotometric 
standards were selected from \markcite{1990ApJ...358..344M}{Massey} \& {Gronwall} (1990) for objects with northern declinations 
and from \markcite{1994PASP..106..566H}{Hamuy} {et~al.} (1994) for southern declinations.  The standards were observed at several 
intervals throughout each night.  In addition to the spectrophotometric standards, giant stars 
covering the spectral range from G0III to M3III, and two super-giants of class M1Iab and M2Iab 
were observed with the same instrumental setup for use as template stars for the kinematic analysis.  
The observed template stars are listed in Table 4.  \\
\clearpage
{
\begin{deluxetable}{lcc}
\tabletypesize{\normalsize}
\setlength{\tabcolsep}{0.1in}
\tablewidth{0pt}
\tablenum{3}
\pagestyle{empty}
\tablecaption{Merger Spectroscopic Observation Log}
\tablecolumns{3}
\tablehead{
\colhead{Merger Name} &
\colhead{Integration Time} &
\colhead{P.A.} \\
\colhead{} &
\colhead{(sec)}&
\colhead{$\circ$}
}
\startdata
\cutinhead{ESI Merger Observations}
UGC 6       &540  &90.0  \\
NGC 34      &1200 &-41.0 \\
NGC 455     &1800 &-30.0 \\
NGC 1210    &300  &-41.0 \\
NGC 1614    &1800 &32.9  \\
AM 0612-373 &1800 &40.0  \\
NGC 2418    &1800 &30.8  \\
NGC 2623    &1800 &64.1  \\
UGC 4635    &1800 &49.8  \\
NGC 2655    &1800 &83.8  \\
NGC 2782    &1800 &90.0  \\
NGC 2914    &1800 &20.5  \\
UGC 5101    &1800 &83.0  \\
NGC 3256    &629  &0.0   \\
Arp 156     &3600 &-61.8 \\
NGC 3597    &1800 &76.7  \\
NGC 3656    &900  &-8.7  \\
NGC 3921    &1800 &29.5  \\
NGC 4004    &1800 &-12.0 \\
NGC 4194    &1800 &-20.0 \\
NGC 4441    &1800 &2.0   \\
AM 1255-430 &2700 &-77.2 \\
NGC 5018    &1140 &90.0  \\
Arp 193     &3600 &-39.3 \\
AM 1419-263 &1800 &69.0  \\
UGC 9829    &1800 &-15.0 \\
NGC 6052    &1800 &71.5  \\
UGC 10607   &1800 &0.0   \\
UGC 10675   &1800 &90.0  \\
AM 2038-382 &1200 &-45.0 \\
AM 2055-425 &1200 &-35.0 \\
NGC 7135    &1800 &0.0   \\
UGC 11905   &1200 &49.5  \\
NGC 7252    &1800 &-60.0 \\
AM 2246-490 &1200 &-5.0  \\
IC 5298     &900  &29.7  \\
NGC 7585    &900  &-70.0   \\
NGC 7727    &900  &90.0  \\
\cutinhead{ESI Elliptical Observations}
NGC 5812    &900  &61.4  \\
\cutinhead{NIRSPEC Merger Observations}
NGC 1614    &900  &32.9  \\
NGC 2623    &850  &64.1  \\
\enddata
\end{deluxetable}

}
\clearpage
{
\begin{deluxetable}{lccc}
\tabletypesize{\small}
\setlength{\tabcolsep}{0.1in}
\tablewidth{0pt}
\tablenum{4}
\pagestyle{empty}
\tablecaption{Template Stars}
\tablecolumns{4}
\tablehead{
\colhead{Name} &
\colhead{R.A.} &
\colhead{Dec.} &
\colhead{Type} \\
}
\startdata
\cutinhead{ESI Observations}
HD 232766         &03$^{h}$ 13$^{m}$ 57$^{s}$ & 54$^{\circ}$ 56$^{'}$ 08$^{''}$ &M1Iab\\
HD 283778         &04$^{h}$ 42$^{m}$ 58$^{s}$ & 27$^{\circ}$ 54$^{'}$ 51$^{''}$ &M0III\\
BD -15 1319       &06$^{h}$ 15$^{m}$ 01$^{s}$ &-15$^{\circ}$ 52$^{'}$ 49$^{''}$ &K5III\\
HD 260158         &06$^{h}$ 36$^{m}$ 14$^{s}$ & 29$^{\circ}$ 31$^{'}$ 15$^{''}$ &K0III\\
HD 50567          &06$^{h}$ 52$^{m}$ 35$^{s}$ &-29$^{\circ}$ 30$^{'}$ 26$^{''}$ &M2-3III\\ 
HD 97646          &11$^{h}$ 14$^{m}$ 02$^{s}$ &-12$^{\circ}$ 47$^{'}$ 38$^{''}$ &K5III\\
HD 99724          &11$^{h}$ 28$^{m}$ 15$^{s}$ &-18$^{\circ}$ 38$^{'}$ 32$^{''}$ &K3III\\
HD 99814          &11$^{h}$ 28$^{m}$ 52$^{s}$ &-16$^{\circ}$ 35$^{'}$ 41$^{''}$ &M3III\\
HD 100059         &11$^{h}$ 30$^{m}$ 43$^{s}$ &-19$^{\circ}$ 53$^{'}$ 46$^{''}$ &K0III\\
HD 100347         &11$^{h}$ 32$^{m}$ 39$^{s}$ &-18$^{\circ}$ 52$^{'}$ 15$^{''}$ &G8III\\
HD 100745         &11$^{h}$ 35$^{m}$ 30$^{s}$ &-19$^{\circ}$ 31$^{'}$ 59$^{''}$ &M0III\\
GSC 02146-01226   &19$^{h}$ 38$^{m}$ 04$^{s}$ & 27$^{\circ}$ 58$^{'}$ 41$^{''}$ &G5III\\
BD +19 4103       &19$^{h}$ 39$^{m}$ 18$^{s}$ & 20$^{\circ}$ 10$^{'}$ 59$^{''}$ &M2Iab\\
HD 332389         &19$^{h}$ 40$^{m}$ 06$^{s}$ & 29$^{\circ}$ 42$^{'}$ 58$^{''}$ &G0III\\
\cutinhead{NIRSPEC Observations}
HD 284318         &04$^{h}$ 18$^{m}$ 59$^{s}$ & 22$^{\circ}$ 08$^{'}$ 28$^{''}$ &K0III\\
\cutinhead{Other CO 2.29 $\micron$ Templates\tablenotemark{a}}
$\alpha$ Orionis  &05$^{h}$ 55$^{m}$ 10$^{s}$ & 07$^{\circ}$ 24$^{'}$ 25$^{''}$ &M1Iab\\
$\lambda$ Dra     &11$^{h}$ 31$^{m}$ 24$^{s}$ & 69$^{\circ}$ 19$^{'}$ 51$^{''}$ &M0III\\
$\alpha$ Boo      &14$^{h}$ 15$^{m}$ 39$^{s}$ & 19$^{\circ}$ 10$^{'}$ 56$^{''}$ &K1.5III\\
Rx Boo            &14$^{h}$ 24$^{m}$ 11$^{s}$ & 25$^{\circ}$ 42$^{'}$ 13$^{''}$ &M7.5III\\
\enddata
\tablecomments{(a) from Wallace \& Hinkle 1996}
\end{deluxetable}

}
\clearpage
\subsection{2.29 $\micron$ CO spectroscopy}
\indent Infrared spectroscopic observations centered on the 2.29 $\micron$ CO feature were obtained 
with the NIRSPEC spectrograph \markcite{1998SPIE.3354..566M}({McLean} {et~al.} 1998) on the W. M. Keck-II 10-meter telescope.  The 
infrared spectrograph contains a 1024$\times$1024 ALADDIN InSb array.  NIRSPEC covers the wavelength range
from 0.95-5.5 $\micron$.  The observations were made in the echelle mode using the 
0{$\arcsec$}.432 $\times$ 24{$\arcsec$} slit.  This gives a resolution of 
{\it R} $\simeq$ 12 km s$^{-1}$ or {\it R} $\simeq$ 25,000.  NIRSPEC contains only a single
echelle grating and a single cross-disperser.  This requires 25 different grating settings to span
the entire 0.95-5.5$\micron$ wavelength range.  The NIRSPEC-7 filter was used, which transmits light 
between 1.839-2.630 $\micron$ and is dispersed over 13 orders. Because
the area of the array is smaller than the area onto which the orders are dispersed, 
only 6-7 non-contiguous orders are projected onto the array at any given time.  
The actual wavelength range of any particular order projected onto the array 
is $\lambda$ $\simeq$ 0.33 $\micron$.
The grating settings were set to the redshifted CO absorption line for the galaxies and standards,
and 2.29 $\micron$ for the template star. The slit Position Angle was placed along the major 
axis of each galaxy. The P.A. was determined using the {\it K}-band data and is listed in Table 3
for each galaxy.  Due to the redshift of the galaxies observed, and the wavelength range of the orders,
the CO feature could not be centered in the middle of the array, but was shifted slightly right of center.  
Since the galaxies were larger than the length of the slit, the observations required nodding to a 
blank area of sky in order to subtract the sky lines and background.  Observations were done 
in an A-B-B-A pattern.  In addition to telluric standards, a K0III star was observed
with the same instrumental setup for use as template star for the kinematic analysis.  
Observations of telluric standards and template star observations were made by nodding along the slit.  
Ne-Ar-Xe-Kr arcs and internal flats were taken after observations of each object and standard
star.

\section{Data Reduction and Analysis}
\subsection{{\it K}-band imaging}
\indent The data set was reduced using IRAF.  The global photometric parameters {\it R$_{\rm eff}$},
$<$$\mu$$_{K}$$>$$_{\rm eff}$, and {\it M$_{K}$} were measured using circular apertures.
The IRAF ELLIPSE package in STSDAS was used to extract the total flux in circular apertures.
Circular apertures were selected over elliptical apertures because they are less sensitive 
to sky-subtraction errors and to match the method used by P99.  \\
\indent The ELLIPSE fitting was conducted by first using IMEXAMINE to find the maximum brightness at the 
center of the galaxy.  The center position was then held fixed.
The radius of each isophote was increased in linear steps of size equal to the
seeing.  The seeing estimates for the galaxies were obtained by measuring the FWHM of high 
signal-to-noise (S/N) stars in 
the field and taking the average value.  Stars in the field were masked and those pixels were ignored in
the isophote fitting and flux measurement.  Only the number of pixels (area in pixels) and the total flux in 
each circular aperture from the circular apertures were used from ELLIPSE.  \\
\indent  The output from ELLIPSE was put into an IDL program which computes the surface brightness, 
S/N and errors at each circular aperture radius.  The program fits a de Vaucouleurs {\it r$^{1/4}$}  
profile to each galaxy using a Levenberg-Marquardt non-linear least squares minimization
technique to achieve a good fit \markcite{1992nrfa.book.....P}({Press} {et~al.} 1992).  All data points were used in the fits.  
The summed total flux in the circular apertures was used to compute 
{\it M$_{K}$}.  \\
\indent It is important to make clear the distinction in terminology between {\it r$_{\rm eff}$} 
and {\it R$_{\rm eff}$}.  The former refers to the effective radius in arcseconds, whereas the latter is 
a metric value in kiloparsecs.  {\it R$_{\rm eff}$} will be used throughout the rest of the paper. 
No corrections to the {\it K}-band data have been made for galactic extinction.  Since the reddening 
at {\it K} is very small, any such corrections, even for objects at low galactic latitude 
would be negligible and well within the errors from standard star calibrations.  Table 5 lists
the extracted photometric properties of the mergers, including 
{\it R}$_{\rm eff}$, $<$$\mu$$_{K}$$>$$_{\rm eff}$, and {\it M$_{K}$}.\\
\clearpage
{
\begin{deluxetable}{lccc}
\tabletypesize{\normalsize}
\setlength{\tabcolsep}{0.1in}
\tablewidth{0pt}
\tablenum{5}
\pagestyle{empty}
\tablecaption{{\it K}-band Photometric Parameters}
\tablecolumns{4}
\tablehead{
\colhead{Merger Name} &
\colhead{Log {\it R$_{\rm eff}$}} &
\colhead{$<$$\mu$$_{K}$$>$$_{eff}$} &
\colhead{{\it M$_{K}$}} \\
\colhead{} &
\colhead{(kpc)} &
\colhead{(mag arcsec$^{-2}$)} &
\colhead{(mag)}\\
}
\startdata
UGC 6         &0.145   &15.18      &-24.01\\
NGC 34        &-0.078  &13.24      &-24.61\\
Arp 230       &0.034   &16.71      &-21.75\\
NGC 455       &0.523   &16.32      &-24.64\\
NGC 828       &0.545   &15.69      &-25.36\\
UGC 2238      &0.151   &14.44      &-24.58\\
NGC 1210      &0.385   &16.64      &-23.72\\
AM 0318-230   &0.561   &16.10      &-25.09\\
NGC 1614      &0.227   &14.92      &-24.74\\
Arp 187       &0.640   &16.33      &-25.25\\
AM 0612-373   &0.673   &16.08      &-25.65\\
NGC 2418      &0.682   &16.40      &-25.31\\
UGC 4079      &0.562   &17.47      &-23.78\\
NGC 2623      &0.120   &14.83      &-24.22\\
UGC 4635      &0.395   &15.73      &-24.71\\
NGC 2655      &0.058   &14.96      &-23.70\\
NGC 2744      &0.536   &18.11      &-22.83\\
NGC 2782      &0.519   &17.04      &-23.83\\
NGC 2914      &0.143   &15.82      &-23.51\\
UGC 5101      &0.028   &12.82      &-25.50\\
AM 0956-282   &0.339   &19.25      &-20.50\\
NGC 3256      &0.254   &14.83      &-24.72\\
NGC 3310      &-0.153  &15.51      &-22.07\\
Arp 156       &0.843   &16.96      &-25.81\\
NGC 3597      &-0.082  &14.21      &-23.72\\
NGC 3656      &0.406   &16.62      &-23.70\\
NGC 3921      &0.538   &16.18      &-25.13\\
NGC 4004      &0.501   &17.90      &-22.89\\
AM 1158-333   &0.150   &16.58      &-22.61\\
NGC 4194      &-0.246  &14.03      &-23.21\\
NGC 4441      &0.186   &16.37      &-22.98\\
UGC 8058      &-0.087  &10.68      &-27.55\\
AM 1255-430   &0.714   &17.11      &-24.93\\
AM 1300-233   &0.641   &16.96      &-24.65\\
NGC 5018      &0.418   &15.33      &-25.15\\
Arp 193       &0.198   &15.10      &-24.40\\
AM 1419-263   &0.557   &16.25      &-24.94\\
UGC 9829      &0.820   &17.68      &-24.96\\
NGC 6052      &0.683   &17.96      &-23.55\\
UGC 10607     &0.202   &14.15      &-25.20\\
UGC 10675     &0.164   &14.66      &-24.80\\
NGC 6598      &0.784   &16.75      &-25.51\\
AM 2038-382   &0.249   &15.12      &-24.70\\
AM 2055-425   &0.320   &14.93      &-25.08\\
NGC 7135      &0.639   &17.47      &-23.95\\
UGC 11905     &0.277   &15.26      &-24.51\\
NGC 7252      &0.403   &15.53      &-24.84\\
AM 2246-490   &0.619   &16.01      &-25.52\\
IC 5298       &0.281   &14.90      &-24.92\\
NGC 7585      &0.648   &16.52      &-24.98\\
NGC 7727      &0.357   &15.86      &-24.23\\
\cutinhead{Elliptical}
NGC 5812      &0.252   &15.52      &-24.08\\

\enddata
\end{deluxetable}

}
\clearpage
\subsection{Optical Spectroscopy}
\indent  Only the orders containing the Ca triplet absorption line at $\lambda$ $\simeq$ 8500 {\AA} 
(order 7) and the Mg$_{2}$ absorption line at $\lambda$ = 5128 {\AA} (orders 11 and/or 12 depending on 
how far the line and continuum are redshifted) were analyzed for the data presented in this paper.  
The spectroscopic data were reduced using IRAF. The spectra were bias-subtracted, and cosmic-rays
were identified and fixed in each image.  A bright spectrophotometric standard was used to trace 
the curved orders before extraction.  The spectra were then extracted in ``strip'' mode with the APALL 
task.  This produced a two-dimensional, rectified spectrum.  Dome flats and calibration arcs were also 
extracted in the same manner.  The data were then reduced in a manner similar to that for long-slit 
spectra.  Each order was divided by the corresponding normalized flat for that order.  
One-dimensional spectra were extracted in an aperture of diameter equivalent to 1.53 {\it h}$^{-1}$$_{75}$ 
kpc for each object.  This aperture size was selected to match the size used in the near-infrared 
Fundamental Plane study by P98 and P99.  The order 
containing the Ca triplet spectra was wavelength calibrated using a Hg-Ne-Xe lamp spectra.  The 
orders containing the Mg$_{2}$ line and continuum were wavelength calibrated using a Cu-Ar lamp spectra.  
The rms of the residuals of the wavelength solutions were 0.09 {\AA} or better.
A sky spectrum was measured at both edges of the slit.  
A 2nd order Legendre polynomial was fit to the background and then subtracted from the spectra.  Next, 
the extracted spectra were corrected to a heliocentric rest velocity.  The spectra were then 
flux-calibrated and continuum normalized by fitting a 5th order spline3 function to the continuum.
The Ca triplet line lies in order 7, which covers the wavelength range from 8117-9366 {\AA}.  This order 
is affected by strong H$_{2}$O telluric absorption longwards of 9000 {\AA}.  Depending on the 
redshift of the object, the Ca triplet lines can lie within this feature.  As a result, the 
spectra were corrected for telluric absorption using the IRAF task TELLURIC.  

\subsubsection{Ca triplet}
\indent The spectra were convolved with a Gaussian equal to the number of pixels 
in one resolution element.  The measurements of the velocity dispersion were carried out using 
a direct template fitting routine.  The direct fitting routine occurs in pixel-space, and is 
based on the methods first described by \markcite{1992MNRAS.254..389R}{Rix} \& {White} (1992).  The advantage to this method is 
that it is straightforward to mask out unwanted features and compute a minimized chi-square 
over an exact wavelength range.  It is more computationally intensive than other techniques 
such as the cross-correlation method \markcite{1979AJ.....84.1511T}({Tonry} \& {Davis} 1979) and the Fourier quotient method 
(e.g. \markcite{1978ApJ...221..731S}{Sargent} {et~al.} (1978)).  However, it has several advantages over these techniques. 
First, there is no need to worry about the ends of the spectrum, which can introduce significant 
high-frequency noise in Fourier space.  Second, absorption features in the spectrum other than those 
under analysis do not interact with each other as they do in Fourier space.  Finally, template mismatch 
errors are significantly reduced in pixel space.\\
\indent The process of pixel-space fitting is simple and straightforward.  The template star
is broadened until the $\chi$$_{\nu}$$^{2}$ of the fit reaches a minimum.  Thus the best-fit 
parameters are those which minimize the $\chi$$_{\nu}$$^{2}$ for the difference between the broadened 
template and galaxy.  Recently, \markcite{2002AJ....124.2607B}{Barth}, {Ho}, \&  {Sargent} (2002) defined an optimal fitting region 
to use for both the \ion{Mg}{1}{\it b} and Ca triplet absorption features.  The same fitting 
region of 8480-8690 {\AA} has been adopted here.  However, unlike Barth, Ho, \& Sargent, who 
assumed a Gaussian profile for the line-of-sight velocity distribution (LOSVD), 
a Gauss-Hermite series polynomial was used to parameterize the Ca triplet line profiles.  
The Gauss-Hermite series is described in detail by \markcite{1993ApJ...407..525V}{van der Marel} \& {Franx} (1993).  The 
Gauss-Hermite function is a modified Gaussian with additional parameters that parameterize 
departures from a Gaussian shape. 
Briefly, the form of the fitting function is:

\begin{equation} {\emph{L}}(\upsilon) = \gamma\frac{\alpha(\omega)}{\sigma}\left[1 + 
{\it h}{\rm_3}{\it H}{\rm_3}(\omega) + 
{\it h}{\rm_4}{\it H}{\rm_4}(\omega)\right] \end{equation}

where $\omega$ $\equiv$ ($\upsilon$ - $\upsilon$$_{\circ}$)/$\sigma$ and

\begin{equation}\alpha(\omega) \equiv \frac{1}{\sqrt[]2\pi} \; e^{-\omega^2/2} \end{equation}
\begin{equation}{\it H}{\rm_3}(\omega) \equiv \frac{1}{\sqrt[]6}(2 \; {\sqrt[]2\omega^3} - 
3 \; {\sqrt[]2\omega}) \end{equation}
\begin{equation}{\it H}{\rm_4}(\omega) \equiv \frac{1}{\sqrt[]24}(4\omega^4 - 12\omega^2 + 3) \end{equation}

When {\it h}$_{\rm 3}$ = {\it h}$_{\rm 4}$ = 0. the Gauss-Hermite series becomes a normal Gaussian profile.  
The fitting function has five parameters: the line strength $\gamma$, which measures the ratio of the 
equivalent width of the galaxy to that of the template star;  the mean recessional 
velocity $\upsilon$$_{\circ}$, the central velocity dispersion $\sigma$$_{\circ}$, the skewness 
{\it h}$_{\rm 3}$, and kurtosis {\it h}$_{\rm 4}$.  Skewness parameterizes the asymmetric 
departures from a Gaussian, while the kurtosis parameterizes the symmetric 
departures from a Gaussian.  Crudely, a positive {\it h}$_{\rm 3}$ shifts the Gaussian to the left, 
a negative {\it h}$_{\rm 3}$ shifts the Gaussian to the right; while a positive {\it h}$_{\rm 4}$ 
makes the peak of the Gaussian more triangular and a negative {\it h}$_{\rm 4}$ squashes the peak
of the Gaussian downwards.  \\
\indent The fitting program uses a Levenberg-Marquardt non-linear least squares minimization 
technique to achieve a good fit \markcite{1992nrfa.book.....P}({Press} {et~al.} 1992).  The five parameters, $\gamma$, 
$\upsilon$$_{\circ}$, $\sigma$, {\it h}$_{\rm 3}$, and {\it h}$_{\rm 4}$ are simultaneously 
fit to the data over the specified wavelength range.  A single stellar population template 
was used in the fitting.  All of the stellar templates listed in Table 4 were fit to each galaxy.
The best fitting template for each galaxy was chosen based on the reduced chi-square and 
rms of the fit. The best-fit template star used for each galaxy is listed in Table 6 along with the results of the fit.  
The parameters listed were extracted using the best fitting template.  \\
\indent The errors shown in Table 6 are not absolute, and are provided more as reasonable estimates.  The error
analysis was conducted by testing various limits on the fitting process.  First, Monte Carlo simulations were
conducted to test the fitting program.  The testing was based on 100 realizations of a template star convolved
with a Gauss-Hermite polynomial of known properties with random noise added.  This altered template star was used
as a ``test galaxy,'' to determine whether the fitting program could recover the input parameters of 
$\gamma$, $\upsilon$$_{\circ}$, $\sigma$$_{\circ}$, {\it h}$_{\rm 3}$, and {\it h}$_{\rm 4}$.
Next, a second template star of identical stellar type was used to recover the input parameters of the ``test galaxy.''
The spread in errors from the Monte Carlo simulations were found to be nearly the same as the fitting errors
determined by the program.  Finally, template mis-match was tested by investigating the spread in the derived parameters
from using different template stars.  Only template stars which produced fits within 2$\times$$\chi$$_{\nu}$$^{2}$ of 
the best-fitting template were used to test each galaxy.  The results show that template mis-match produced the largest
errors in the fitting.  The errors listed in Table 6 for each parameter are the standard deviations of the derived parameters
for the range of template stars used to test the template mis-match.  The spectrum for each galaxy is shown in Appendix A.
The solid line is the galaxy spectrum and the overplotted dashed line is the convolved best-fitting template.
\clearpage
{
\begin{deluxetable}{lcccccc}
\tabletypesize{\normalsize}
\setlength{\tabcolsep}{0.1in}
\tablewidth{0pt}
\tablenum{6}
\pagestyle{empty}
\rotate
\tablecaption{Spectroscopic Parameters}
\tablecolumns{7}
\tablehead{
\colhead{Merger Name} &
\colhead{{$\sigma$$_{\circ}$}} &
\colhead{{\it V$_{\odot}$}} &
\colhead{$\gamma$} &
\colhead{{\it h}$_{3}$} &
\colhead{{\it h}$_{4}$} &
\colhead{Best-fit Template} \\
\colhead{} &
\colhead{(km s$^{-1}$)} &
\colhead{(km s$^{-1}$)} &
\colhead{} &
\colhead{} &
\colhead{} &
\colhead{star/type}
}
\startdata
\cutinhead{ESI Observations}
UGC 6            &220  $\pm$ 10  &6579  $\pm$  1  &0.906 $\pm$ 0.005  &-0.013  $\pm$ 0.009  & 0.185 $\pm$ 0.018   &HD 332389 G0III\\
NGC 34           &201  $\pm$  8  &5881  $\pm$  2  &1.054 $\pm$ 0.014  & 0.000  $\pm$ 0.006  &-0.079 $\pm$ 0.030   &HD 332389 G0III\\
NGC 455          &234  $\pm$  7  &5827  $\pm$  1  &0.970 $\pm$ 0.012  &-0.005  $\pm$ 0.004  & 0.069 $\pm$ 0.026   &HD 100059 K0III\\
NGC 1210         &247  $\pm$  6  &3878  $\pm$  7  &0.953 $\pm$ 0.011  & 0.052  $\pm$ 0.016  & 0.116 $\pm$ 0.021   &HD 283778 M0III\\
NGC 1614         &146  $\pm$ 12  &4769  $\pm$  1  &0.921 $\pm$ 0.006  &-0.031  $\pm$ 0.006  & 0.136 $\pm$ 0.020   &HD 332389 G0III\\
AM 0612-373      &303  $\pm$  8  &9721  $\pm$  2  &0.974 $\pm$ 0.003  &-0.047  $\pm$ 0.009  & 0.101 $\pm$ 0.010   &HD 99724  K3III\\
Arp 165          &288  $\pm$ 10  &5037  $\pm$  1  &0.962 $\pm$ 0.009  & 0.017  $\pm$ 0.003  & 0.091 $\pm$ 0.021   &HD 100347 G8III\\
NGC 2623         &191  $\pm$  7  &5549  $\pm$  1  &0.957 $\pm$ 0.008  &-0.014  $\pm$ 0.006  & 0.106 $\pm$ 0.020   &HD 100347 G8III\\
UGC 4635         &251  $\pm$  7  &8722  $\pm$  1  &0.993 $\pm$ 0.009  & 0.002  $\pm$ 0.003  & 0.056 $\pm$ 0.019   &HD 100347 G8III\\
NGC 2655         &169  $\pm$ 11  &1400  $\pm$  1  &0.995 $\pm$ 0.004  & 0.013  $\pm$ 0.004  & 0.043 $\pm$ 0.011   &HD 100347 G8III\\
NGC 2782         &196  $\pm$  8  &2543  $\pm$  2  &0.946 $\pm$ 0.007  &-0.003  $\pm$ 0.009  & 0.135 $\pm$ 0.019   &HD 100347 G8III\\
NGC 2914          &186  $\pm$  4  &3159  $\pm$  1  &1.010 $\pm$ 0.019  &-0.001  $\pm$ 0.005  & 0.014 $\pm$ 0.034   &HD 100059 K0III\\
UGC 5101         &287  $\pm$ 11  &11802 $\pm$  2  &0.983 $\pm$ 0.009  & 0.000  $\pm$ 0.006  & 0.096 $\pm$ 0.021   &HD 100347 G8III\\
NGC 3256         &241  $\pm$ 16  &2795  $\pm$  8  &0.849 $\pm$ 0.003  &-0.035  $\pm$ 0.030  & 0.348 $\pm$ 0.010   &HD 100347 G8III\\
Arp 156          &288  $\pm$  8  &10738 $\pm$  4  &0.994 $\pm$ 0.006  & 0.162  $\pm$ 0.007  & 0.071 $\pm$ 0.013   &HD 260158 K0III\\
NGC 3597         &174  $\pm$  9  &3500  $\pm$  1  &0.913 $\pm$ 0.009  &-0.047  $\pm$ 0.007  & 0.185 $\pm$ 0.024   &HD 100347 G8III\\
NGC 3656         &132  $\pm$ 18  &2890  $\pm$ 11  &1.035 $\pm$ 0.016  & 0.108  $\pm$ 0.045  &-0.057 $\pm$ 0.032   &HD 332389 G0III\\
NGC 3921         &222  $\pm$  5  &5896  $\pm$  1  &1.010 $\pm$ 0.014  &-0.115  $\pm$ 0.007  & 0.011 $\pm$ 0.027   &HD 100347 G8III\\
NGC 4004         & 33  $\pm$  2  &3368  $\pm$  1  &0.998 $\pm$ 0.024  & 0.001  $\pm$ 0.019  & 0.014 $\pm$ 0.041   &HD 100347 G8III\\
NGC 4194         &116  $\pm$  7  &2501  $\pm$  1  &0.891 $\pm$ 0.004  &-0.086  $\pm$ 0.008  & 0.247 $\pm$ 0.014   &HD 100347 G8III\\
NGC 4441         &139  $\pm$  6  &2722  $\pm$  1  &0.923 $\pm$ 0.012  &-0.060  $\pm$ 0.009  & 0.172 $\pm$ 0.029   &HD 100347 G8III\\
AM 1255-430      &243  $\pm$  3  &8984  $\pm$  2  &1.017 $\pm$ 0.008  &-0.170  $\pm$ 0.010  & 0.039 $\pm$ 0.016   &HD 99724  K3III\\
NGC 5018         &222  $\pm$  3  &2816  $\pm$  1  &0.997 $\pm$ 0.003  &-0.008  $\pm$ 0.006  & 0.035 $\pm$ 0.007   &HD 100059 K0III\\
Arp 193          &172  $\pm$  8  &6985  $\pm$  3  &1.001 $\pm$ 0.008  &-0.032  $\pm$ 0.008  & 0.005 $\pm$ 0.020   &HD 332389 G0III\\
AM 1419-263      &260  $\pm$  6  &6758  $\pm$  1  &0.978 $\pm$ 0.005  & 0.021  $\pm$ 0.003  & 0.070 $\pm$ 0.011   &HD 100059 K0III\\
UGC 9829         &134  $\pm$  4  &8465  $\pm$  1  &0.980 $\pm$ 0.009  &-0.013  $\pm$ 0.007  & 0.074 $\pm$ 0.018   &HD 100059 K0III\\
NGC 6052         & 80  $\pm$  5  &4739  $\pm$  1  &0.881 $\pm$ 0.005  &-0.016  $\pm$ 0.007  & 0.250 $\pm$ 0.017   &HD 100347 G8III\\
UGC 10607        &211  $\pm$  5  &10376 $\pm$  2  &1.002 $\pm$ 0.010  &-0.045  $\pm$ 0.008  & 0.037 $\pm$ 0.021   &HD 100347 G8III\\
UGC 10675        &177  $\pm$  6  &10179 $\pm$  2  &0.971 $\pm$ 0.005  & 0.034  $\pm$ 0.005  & 0.106 $\pm$ 0.013   &HD 100347 G8III\\
AM 2038-382      &257  $\pm$  8  &6092  $\pm$  2  &0.992 $\pm$ 0.012  &-0.049  $\pm$ 0.006  & 0.040 $\pm$ 0.026   &HD 100347 G8III\\
AM 2055-425      &185  $\pm$  6  &12890 $\pm$  2  &1.054 $\pm$ 0.009  &-0.007  $\pm$ 0.007  &-0.085 $\pm$ 0.020   &HD 332389 G0III\\
NGC 7135         &277  $\pm$  9  &2644  $\pm$  1  &0.983 $\pm$ 0.010  & 0.000  $\pm$ 0.003  & 0.070 $\pm$ 0.022   &HD 100347 G8III\\
UGC 11905        &222  $\pm$  9  &7456  $\pm$  1  &0.957 $\pm$ 0.009  &-0.014  $\pm$ 0.014  & 0.110 $\pm$ 0.022   &HD 100347 G8III\\
NGC 7252         &166  $\pm$  5  &4792  $\pm$  1  &0.968 $\pm$ 0.015  &-0.023  $\pm$ 0.007  & 0.076 $\pm$ 0.030   &HD 100347 G8III\\
AM 2246-490      &267  $\pm$  7  &12901 $\pm$  2  &0.967 $\pm$ 0.007  &-0.060  $\pm$ 0.008  & 0.103 $\pm$ 0.016   &HD 100347 G8III\\
IC 5298          &193  $\pm$  6  &8221  $\pm$  1  &0.961 $\pm$ 0.006  &-0.065  $\pm$ 0.004  & 0.118 $\pm$ 0.015   &HD 100347 G8III\\
NGC 7585         &211  $\pm$  4  &3539  $\pm$  1  &0.965 $\pm$ 0.003  &-0.021  $\pm$ 0.009  & 0.085 $\pm$ 0.006   &HD 100059 K0III\\
NGC 7727         &231  $\pm$  5  &1868  $\pm$  2  &0.968 $\pm$ 0.003  & 0.017  $\pm$ 0.005  & 0.089 $\pm$ 0.008   &HD 99724  K3III\\
\cutinhead{Elliptical}
NGC 5812         &241  $\pm$  4  &1970  $\pm$  2  &0.967 $\pm$ 0.003  & 0.000  $\pm$ 0.006  & 0.099 $\pm$ 0.005   &HD 100059 K0III\\
\cutinhead{NIRSPEC Observations}
NGC 1614         &143  $\pm$ 12  &4779  $\pm$  5  &1.068 $\pm$ 0.063  &-0.116  $\pm$ 0.024  & 0.065 $\pm$ 0.135   &Rx Boo M7.5III\\
NGC 2623         &139  $\pm$ 19  &5555  $\pm$ 17  &1.155 $\pm$ 0.086  & 0.078  $\pm$ 0.021  &-0.186 $\pm$ 0.164   &$\lambda$ Dra M3III\\
\enddata
\end{deluxetable}

}
\clearpage
\subsubsection{Mg$_{2}$}
\indent The Mg$_{2}$ absorption feature at 5178 {\AA} was measured using the index bandpass and 
pseudo-continua defined by \markcite{1994ApJS...94..687W}{Worthey} {et~al.} (1994).  The index bandpass is given as 
5154.125-5196.625 {\AA}.  The blue pseudo-continuum is 4895.125-4957.625 and the red 
pseudo-continuum is 5301.125-5366.125.  The index is measured in units of magnitude.  
This bandpass is defined as part of the Lick/IDS system.  It is defined
by the use of the Lick Observatory image dissector scanner (IDS), which has an 
8.4 {\AA} resolution at that wavelength range.  The spectra observed using the Lick IDS spectrograph were not flux-calibrated, 
or corrected for instrument response.  The indices were measured on the original 
spectra \markcite{1985ApJS...57..711F}({Faber} {et~al.} 1985).  A line representing the pseudo-continuum was drawn between the midpoints 
of the two sidebands and the flux difference between this line and the central bandpass flux 
determines the index.  In order to convert to the Lick/IDS system from another spectrograph, it is 
necessary to first convolve the spectra to the same resolution and then calibrate the indices by 
observing ``standards,'' which are any stars previously observed on the Lick IDS spectrograph.\\
\indent The Mg$_{2}$ magnitudes listed in Table 7 are raw magnitudes because they have not been 
fully converted to the Lick/IDS system.  The spectra were convolved with a Gaussian to smooth the resolution 
to 8.4 {\AA}, but unfortunately, no stars previously observed with the Lick IDS spectrograph 
were observed with ESI.  Furthermore, the prescription for data reduction as defined by 
Faber et al. produced problems with the ESI data.  Measurements made on the raw ESI data produced 
spurious and often negative magnitudes (even though the Mg$_{2}$ absorption feature was present). 
This is due to the shape of the spectral response of ESI in those orders.
Instead, the spectra were flux calibrated in order to remove the spectral 
response of the instrument. \markcite{1990ApJ...362..503B}{Brodie} \& {Huchra} (1990) compared results between flux calibrated
and non flux calibrated indices and found a negligible difference.   A comparison of 
the Mg$_{2}$ index for NGC 5812 extracted from the literature by P99 (and calibrated onto the Lick system)
and measured with ESI, using the same aperture size, shows a difference of $\simeq$ 0.02 mag.\\
\clearpage
{
\begin{deluxetable}{lc}
\tabletypesize{\normalsize}
\setlength{\tabcolsep}{0.1in}
\tablewidth{0pt}
\tablenum{7}
\pagestyle{empty}
\tablecaption{Mg$_{2}$ Indices}
\tablecolumns{2}
\tablehead{
\colhead{Merger Name} &
\colhead{Mg$_{2}$} \\
\colhead{} &
\colhead{(mag)\tablenotemark{a}}
}
\startdata
UGC 6         &0.06  \\
NGC 34        &0.04  \\
NGC 455       &0.22  \\
NGC 1210      &0.34  \\
NGC 1614      &0.06  \\
AM 0612-373   &0.21  \\
NGC 2418      &0.30  \\
NGC 2623      &0.08  \\
UGC 4635      &0.29  \\
NGC 2655      &0.19  \\
NGC 2782      &0.10  \\
NGC 2914      &0.26  \\
UGC 5101      &0.13  \\
NGC 3256      &0.03  \\
Arp 156       &0.26  \\
NGC 3597      &0.06  \\
NGC 3656      &0.12  \\
NGC 3921      &0.02  \\
NGC 4004      &0.06  \\
NGC 4194      &0.06  \\
NGC 4441      &0.08  \\
AM 1255-430   &0.13  \\
NGC 5018      &0.17  \\
Arp 193       &0.11  \\
AM 1419-263   &0.26  \\
UGC 9829      &0.20  \\
NGC 6052      &0.08  \\
UGC 10607     &0.14  \\
UGC 10675     &0.04  \\
AM 2038-382   &0.06  \\
AM 2055-425   &0.04  \\
NGC 7135      &0.29  \\
UGC 11905     &0.10  \\
NGC 7252      &0.09  \\
AM 2246-490   &0.05  \\
IC 5298       &0.13  \\
NGC 7585      &0.21  \\
NGC 7727      &0.28  \\
\cutinhead{Elliptical}
NGC 5812      &0.29  \\

\enddata
\tablecomments{(a) Raw Mg$_{2}$ indices.  They have not been fully calibrated to the Lick/IDS system.}
\end{deluxetable}

}
\clearpage
\subsection{2.29 $\micron$ CO spectroscopy}
\indent  Only the order containing the CO absorption feature was analyzed for the data presented 
in this paper.  The spectroscopic data were reduced using IRAF.  A bad-pixel mask was created using 
one of the dark frames and used to mask bad pixels and cosmetic defects on the array.  The data were 
then pair-subtracted to remove background night-sky lines.   Bright standard stars were used to 
trace the curved orders before extraction.  The spectra were then extracted in ``strip'' mode with 
the APALL task.  This produced a two-dimensional, spectrum.  Flats and calibration arcs were also 
extracted in the same manner.  The data were then reduced in a manner similar to that for long-slit 
spectra.  Each order was divided by the corresponding normalized flat for that order.\\  
\indent In addition to a slight curve present in the orders, the spectral lines are also curved along
both the spatial and dispersion axes.  The orders were straightened in IRAF by measuring 
and computing the curvature of the Ne-Ar-Xe-Kr arc lines in the arc calibration frames and applying the
results to the science data frames. 
The spectra were then extracted in an aperture of diameter equivalent to 1.53 {\it h}$^{-1}$$_{75}$ kpc 
for each object.  This aperture size was selected to match the size used in the near-infrared 
Fundamental Plane study by P98 and P99.   The slit width used was 0{$\arcsec$}.432, which
is very close to the size used for the ESI data.  The order was then wavelength calibrated using the Ne-Ar-Xe-Kr arc.
A 2nd order Legendre polynomial was fit to the background and then subtracted 
from the spectra in order to remove any residual background lines that were not properly pair-subtracted.
Next, the extracted spectra were corrected to a heliocentric rest velocity.  The spectra were
then divided by an A0V telluric standard.  This removes telluric atmospheric absorption features which
can be quite strong in the near-infrared. The spectra were not flux calibrated.  The continuum was then 
normalized by fitting a 1st-order Legendre polynomial to the blue-ward side of the CO feature.
This was done because there is no continuum red-ward of the CO feature.\\
\indent The data analysis was identical the methods described under the ESI optical spectroscopy 
section above. The wavelength of the analysis was restricted to a range of 2.2775-2.3005 $\micron$.
This was selected because it was found to be the largest wavelength range in common between the galaxies
observed and the standard stars.  The K0III template star observed with NIRSPEC 
proved to be inadequate in the fitting routine.  The CO feature in the template star was too 
shallow compared to the depth of feature in the target galaxies.  As a result additional template 
stars were taken from the high-resolution {\it K}-band spectral atlas of ordinary cool stars by 
\markcite{1996ApJS..107..312W}{Wallace} \& {Hinkle} (1996). Their data was obtained with a Fourier transform spectrograph at the 
Kitt Peak Mayall 4-meter telescope.  The resolution was {\it R} $\geq$ 45,000.  The data were then 
smoothed to the resolution of NIRSPEC by convolving them with a Gaussian.  Table 4 lists the template 
stars used from Wallace \& Hinkle.  The best fit template star used for the CO absorption measurements 
are listed in Table 6 for the two galaxies observed.  The plotted spectrum for each galaxy is shown in Appendix A.

\section{Results}
\subsection{The Near-Infrared Fundamental Plane}
\indent Figure 1 shows the {\it K}-band Fundamental Plane of elliptical galaxies with the mergers overplotted.
The dotted line is the Fundamental Plane derived from P98.
The mergers are plotted in this figure using the photometric data from Table 5 and the kinematic data from Table 6.
The filled circles are ``normal'' mergers, the open circles are the LIRG/ULIRG mergers, the open diamonds
are shell ellipticals, and the triangle is NGC 5812, the elliptical galaxy common to both this study
and P98.  NGC 1614 and NGC 2623, the two LIRGs with Ca triplet and CO velocity dispersions
are plotted twice.  The Ca triplet and CO derived data points are connected by a solid line to make a comparison easier.
The five-point stars are the data points for NGC 1614 and NGC 2623 using the CO velocity 
dispersion.  The ``x'' symbols in the plot are ellipticals from P99 which were used to derive the Fundamental 
Plane in P98.  While P98 used early-type galaxies, including S0's and some peculiar E's, to construct 
the fundamental plane, only galaxies classified as ``pure'' ellipticals are plotted from the P99 sample.  
They are plotted in addition to the best-fit line in order to show the scatter present among the 
ellipticals and to compare that with the scatter among the mergers.\\
\clearpage
{
\begin{figure}
\plotone{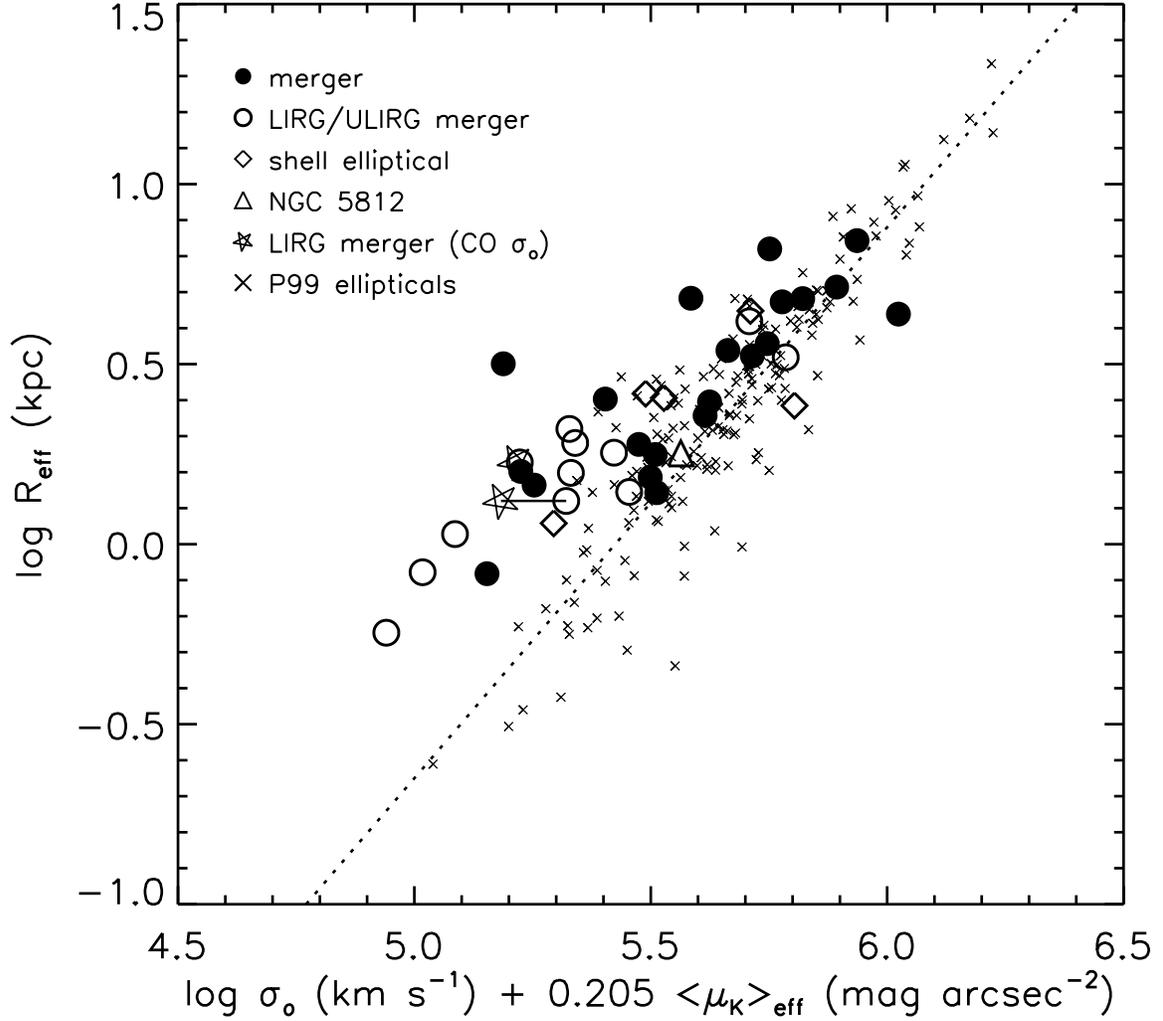}
\caption{The {\it K}-band Fundamental Plane of Elliptical 
Galaxies from P98.  The dotted line is the Fundamental Plane, log {\it R}$_{\rm eff}$ = 
1.53 log $\sigma$$_{\circ}$ + 0.314 $<$$\mu$$_{K}$$>$$_{\rm eff}$ - 8.30.
The filled circles are normal mergers, the open circles are the LIRG/ULIRG mergers, the open diamonds are 
the shell ellipticals, and the five-point stars are the LIRGs NGC 1614 and NGC 2623 plotted using the 
CO derived $\sigma$$_{\circ}$.  Their points are connected by a solid line to the Ca triplet derived
values to make a comparison easier.  The open triangle is the ``control'' sample elliptical, NGC 5812.
The Spearman Rank coefficient indicates that the mergers show a strong correlation between the Fundamental Plane 
parameters at a significance of 0.001.  The slope of the mergers on the Fundamental Plane is 0.84 with a scatter of 0.135 dex.}
\end{figure}
}
\clearpage
\indent In general, a majority of the mergers appear to lie on the Fundamental Plane or within the same scatter of 
the ellipticals from P99.  The mergers show evidence of a strong correlation among
the parameters of the Fundamental Plane.  The Spearman Rank Correlation was used to test this statistically.
This test was selected because it is non-parametric, it makes no assumption about the shape of the distribution.  
The only assumption made is that the distribution is continuous.
The Spearman Rank Correlation coefficient indicates a strong correlation of the parameters at significance level of
0.001.  The scatter of the mergers is only 0.135 dex, compared
with 0.096 dex for the ellipticals from P98.  However, the slope of the merger line is
0.84, which differs from the slope of 1.53 for the {\it K}-band Fundamental Plane.  Visually, the mergers in 
Figure 1 appear to show evidence of a ``tail''-like feature at values of smaller {\it R}$_{\rm eff}$,
brighter $<$$\mu$$_{K}$$>$$_{\rm eff}$ and smaller $\sigma$$_{\circ}$.  Somewhere around
20$\%$ of the mergers appear to lie within this offset region.  No mergers appear to lie beyond
the scatter of the P99 ellipticals to the right of the Fundamental Plane.  If the offset of mergers from the Fundamental
Plane were due purely to random scatter and/or a poor correlation among the parameters, then there should be no
preferred direction or structure.  Two possible explanations for the offset are an evolutionary difference, suggesting these
objects will eventually reach the Fundamental Plane, or these mergers are somehow physically different, and may
never reach the Fundamental Plane.  \markcite{2004A&A...423..833M}{Michard} \& {Prugniel} (2004) conducted a study of peculiar ellipticals taken from
a catalog of early-type galaxies \markcite{1996A&A...309..749P}({Prugniel} \& {Simien} 1996).    They found the peculiar ellipticals systematically 
deviated from both the Fundamental Plane and Faber-Jackson relation.  The primary reason for the deviation
is the presence of a younger population of stars.  The peculiar ellipticals were found to be brighter than expected
and have smaller Mg$_{2}$ indices as compared with non-peculiar or ``normal'' ellipticals.\\
\indent While photometric properties such as {\it R}$_{\rm eff}$ and  $<$$\mu$$_{K}$$>$$_{\rm eff}$ can change over time as a result of
star-formation and stellar evolution, the central velocity dispersion does not.
Merger simulations, i.e. \markcite{1988ApJ...331..699B,1992ApJ...393..484B}({Barnes} 1988, 1992), suggest that once a single nucleus is formed
the central velocity dispersion becomes fixed.  All the mergers in the sample were selected based on the presence of single nucleus,
therefore, their observed $\sigma$$_{\circ}$ is unlikely to change.  The mergers offset from the Fundamental Plane
lie in a region where a lower $\sigma$$_{\circ}$ {\it may} contribute partially or solely to the difference. Therefore, it
is important to test whether the dispersions of the mergers are consistent with elliptical galaxies.  If they are, then
the offset is the result of photometric differences which can change over time.

\subsubsection{Kinematic versus Photometric Differences} 
\indent Figure 2 shows a rotated view of the Fundamental Plane, with the photometric and kinematic observables separated.  The symbols and
overplotted dotted line are the same as Figure 1.  This projection of the Fundamental Plane shows the offset is more the result of photometric 
differences than kinematic ones.  Only one merger, NGC 4004, lies outside of the range of the P99 central velocity dispersions.
Furthermore, the mergers appears to be skewed towards larger values of $\sigma$$_{\circ}$ relative to the P99 sample.\\
\clearpage
{
\begin{figure}
\plotone{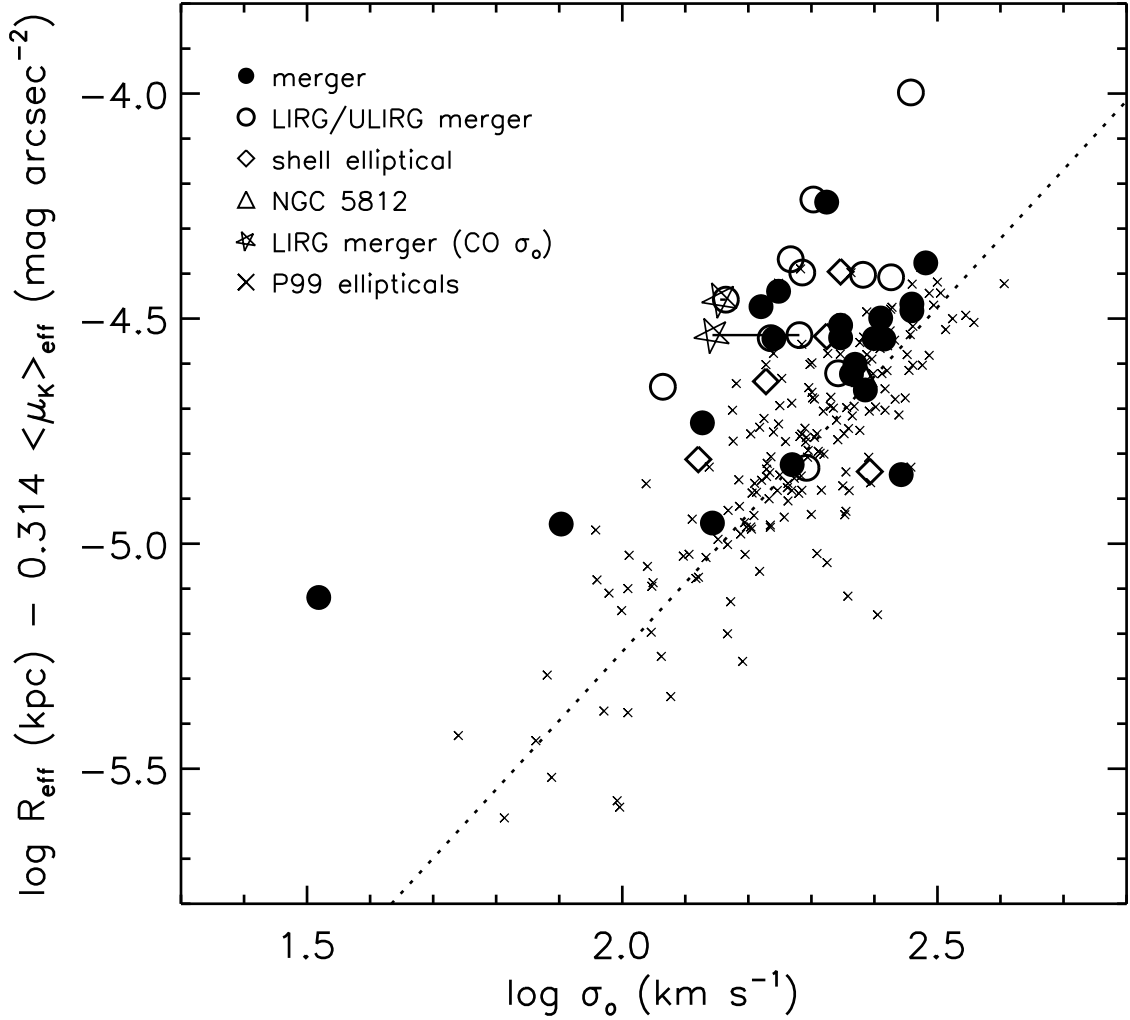}
\caption{Figure 2 shows the Fundamental Plane projected so that the photometrically and kinematically observed properties
are separated.  The symbols and dotted line are otherwise the same as Figure 1.}
\end{figure}
}
\clearpage
\indent A more quantitative analysis was conducted using a Kolmogorov-Smirnov (K-S) two-tailed test to
compare the mergers and each sub-sample with other galaxy samples.  The K-S test probes the null hypothesis
that two two distributions in question arise from the same parent population.  It is a non-parametric test,
it makes no assumptions about the form of the parent distribution.  The only assumption is that the two distributions
are continuous.  A K-S test comparing the $\sigma$$_{\circ}$ of the merger sample and the P99 sample of elliptical galaxies shows
the null hypothesis cannot be rejected.  That is, they appear to arise from the same parent population.
A comparison of the log {\it R}$_{\rm eff}$ between the mergers and P99 ellipticals produces
the same results. Finally, a comparison of the $<$$\mu$$_{K}$$>$$_{\rm eff}$ between the mergers and P99 ellipticals rejects the null hypothesis
at the 0.001 confidence level.  These statistical results in conjunction with Figure 2 suggest that
the difference in slope between the mergers and P99 ellipticals is more likely the result of photometric, not
kinematic differences.  The presence of a preferred direction in the offset of some mergers from the Fundamental Plane
is also likely due to differences in the effective radius and higher surface brightness, rather than a low $\sigma$$_{\circ}$.

\subsubsection{Do Mergers form Massive Ellipticals?}
\indent If mergers show a strong correlation among the Fundamental Plane parameters, an important question to
ask is what type of elliptical they are likely to form.  This question has been touched upon before in the literature. 
 \markcite{1982ApJ...252..455S,1996AJ....111..109S}{Schweizer} (1982, 1996) points out that the optical luminosity of NGC 3921 and NGC 7252 are
high enough to fall within the range of luminous giant ellipticals (gE's).  The results from Paper I suggest that most of the mergers,
including LiRG and ULIRG types, are luminous enough at {\it K}-band to form ellipticals which are $\ge$ {\it L}$^{*}$.
However, \markcite{2001ApJ...563..527G}{Genzel} {et~al.} (2001) argue, based on the 
central velocity dispersions derived from the CO absorption line at 1.63 and 2.29 $\micron$, that ULIRG mergers will form 
only $\sim$ {\it L}$^{*}$ intermediate-mass ellipticals and cannot form gE's.  Citing the earlier CO studies of  
\markcite{1998ApJ...497..163S}{Shier} \& {Fischer} (1998) and \markcite{1999MNRAS.309..585J}{James} {et~al.} (1999), they further suggest that LIRG mergers will form 
sub-{\it L}$^{*}$ intermediate-mass ellipticals.  \markcite{1999MNRAS.309..585J}{James} {et~al.} (1999) note for their sample
that a K-S test shows that the central velocity dispersions of their merger sample are consistent with {\it both} spiral bulges and elliptical galaxies.
These results are in stark contrast to predictions that LIRG/ULIRG mergers might be analogous to the objects which formed present-day gE's
at z $>$ 1 \markcite{1992ApJ...390L..53K}({Kormendy} \& {Sanders} 1992).  One reason ULIRGs could be potential analogues to these objects for is the large quantities of molecular gas 
in these objects which provide enough fuel to trigger strong star-formation.\\
\indent The kinematic data presented here provides an opportunity to test whether or not the luminosity and kinematics are both  
consistent with the suggestion that mergers can produce luminous giant ellipticals.  Kinematic data were taken from the literature
for samples of gE's, intermediate-mass ellipticals, as well as spiral bulges, to test whether the kinematics and luminosities of
the mergers produce consistent results or show the same disparities as earlier studies.
A comparison with the bulges of spiral galaxies was also made because bulges are known to lie on the Fundamental Plane
and posses stellar profiles similar to elliptical galaxies.  The survival of a bulge during a spiral-spiral merger could produce elliptical-like 
properties, including an {\it r}$^{1/4}$ light profile within the inner several kpc.
The sample of giant E's and intermediate-mass ellipticals were taken from \markcite{1992ApJ...399..462B}{Bender} {et~al.} (1992) (hereafter BBF92).
Only ellipticals designated as ``pure-E'''s were used from BBF92.  They divided their larger sample of ellipticals based on absolute luminosity 
and morphology:
Giant elliptical were defined as {\it  M}$_{B}$ $\leq$ -19.62, intermediate-mass ellipticals were defined 
as -19.62 $<$ {\it M}$_{B}$ $\leq$ -17.62 (for {\it H}$_{\circ}$ = 75 km s$^{-1}$ Mpc$^{-1}$). In order to 
reduce possible errors introduced by the comparison of different aperture size, the $\sigma$$_{\circ}$ 
of the ellipticals were corrected to the aperture diameter of 1.53 kpc used by P99. \\
\indent  The $\sigma$$_{\circ}$ for 98 spiral bulges were taken from \markcite{1979ApJ...234...68W}{Whitmore}, {Schechter}, \&  {Kirshner} (1979),
\markcite{1998A&AS..133..317H}{H{\' e}raudeau} \& {Simien} (1998), \markcite{1999A&AS..136..509H}{H{\' e}raudeau} {et~al.} (1999), \markcite{2002MNRAS.335..741F}{Falc{\' o}n-Barroso},  {Peletier}, \& {Balcells} (2002) and 
\markcite{2003A&A...405..455F}{Falc{\' o}n-Barroso} {et~al.} (2003).  No aperture corrections were applied to the $\sigma$$_{\circ}$ of the bulges because it is not clear 
that $\sigma$$_{\circ}$ changes with radius in the same fashion as elliptical galaxies.  Aperture 
sizes were on average a few arcseconds in size.  The sample selection was limited to spirals with 
{\it cz} $\leq$ 4000 km s$^{-1}$ and  {\it M}$_{B}$ $\leq$ -18 
(for {\it H}$_{\circ}$ = 75 km s$^{-1}$ Mpc$^{-1}$).  The late-type galaxies range from Sa through Scd.\\
\indent The results indicate that the hypothesis that mergers and spiral bulges arise from the same parent
population can be rejected at the 0.001 confidence level.  The same results occur for each of the 
merger sub-samples, except the shell ellipticals.  The hypothesis there can be rejected at the 0.05 confidence level.
When the merger sample as a whole is compared with gE's, the hypothesis that the two samples
arise from the same parent population can be rejected at the 0.1 confidence level, which is rather weak.
The hypothesis cannot be rejected for the merger sample as a whole when compared with the intermediate-mass
ellipticals.  \\
\indent When the each of the merger sub-samples are compared separately with gE's and intermediate-mass ellipticals,
the results change somewhat.  Table 8 shows the results of the K-S test for the mergers, merger sub-samples,
as well as the LIRG/ULIRG mergers from the studies noted above, and mergers from \markcite{1986ApJ...310..605L}{Lake} \& {Dressler} (1986).
They were all tested against gE's, intermediate-mass ellipticals, and spiral bulges.
The $\sigma$$_{\circ}$  of the mergers from \markcite{1986ApJ...310..605L}{Lake} \& {Dressler} (1986) include those derived from both
the Calcium triplet and Mg {\it Ib} stellar absorption lines. The velocity dispersions derived from
the Calcium triplet took precedence in the analysis, although the dispersions derived from both lines were similar.
Figure 3 illustrates the results in histogram form.  The bin size is 25 km s$^{-1}$.  The data are plotted as a fraction of
the total number for each sample.\\
\clearpage
{
\begin{deluxetable}{lccc}
\tabletypesize{\normalsize}
\setlength{\tabcolsep}{0.1in}
\tablewidth{0pt}
\tablenum{8}
\pagestyle{empty}
\rotate
\tablecaption{Confidence Levels for the Rejection of the Null Hypothesis that
Two Samples Share the Same Parent Population}
\tablecolumns{5}
\tablehead{
\colhead{Sample} &
\colhead{Giant Ellipticals} &
\colhead{Intermediate-mass Ellipticals} &
\colhead{Bulges} \\
\colhead{}&
\colhead{Confidence} &
\colhead{Confidence} &
\colhead{Confidence} \\
\colhead{} &
\colhead{Level}&
\colhead{Level}&
\colhead{Level}
}
\startdata
All Mergers                    &0.01     &\nodata   &0.001  \\
Normal Mergers                 &\nodata  &0.1       &0.001  \\
ULIRG/LIRG Mergers             &0.1      &\nodata   &0.001  \\
Shell Ellipticals              &\nodata  &\nodata   &0.05   \\
Previous ULIRG/LIRGs studies   &0.001    &\nodata   &0.001   \\
Mergers from Lake \& Dressler  &0.05     &0.1       &0.001    \\
\enddata
\end{deluxetable}

}
\clearpage
{
\begin{figure}
\epsscale{0.95}
\plotone{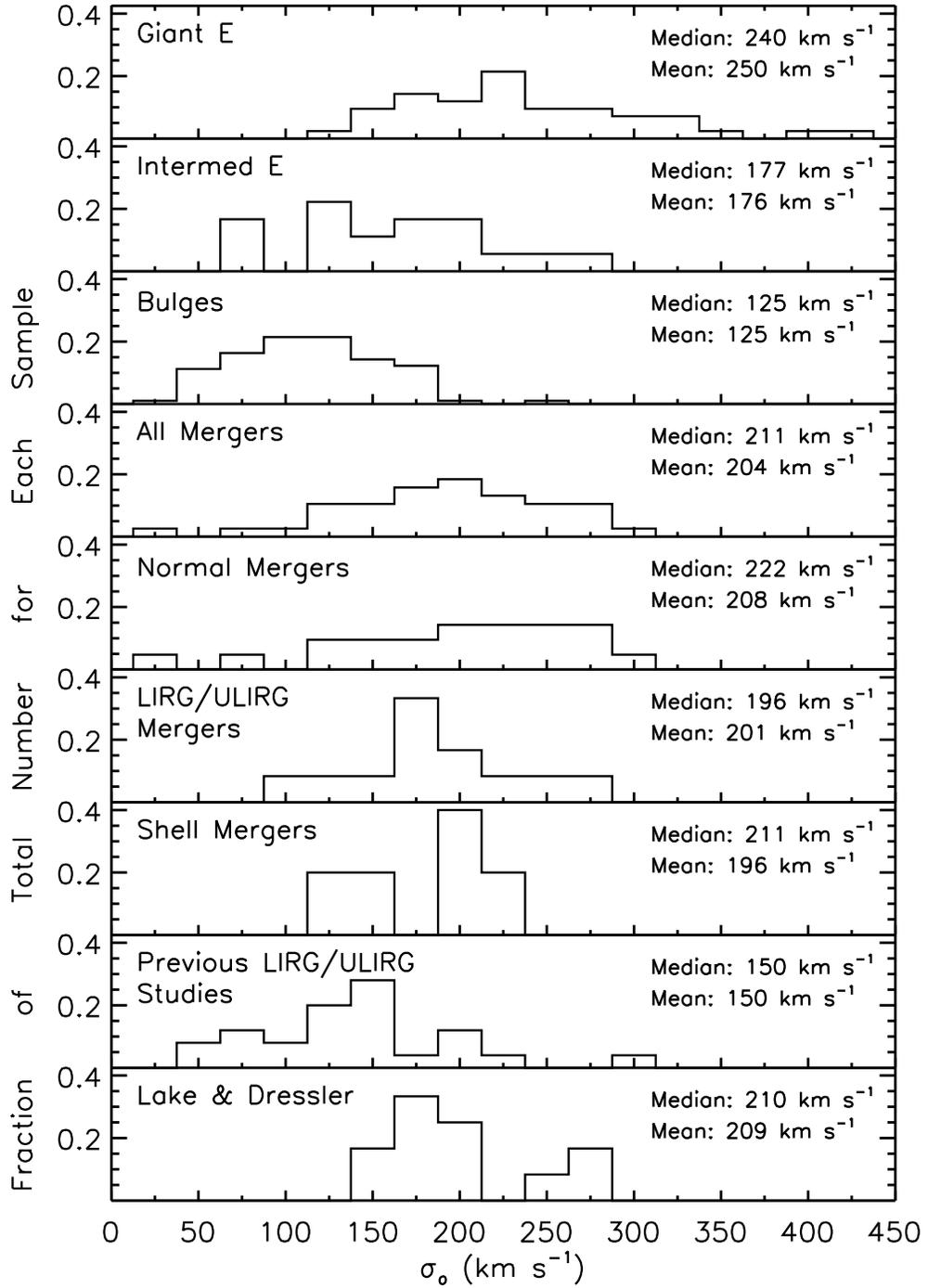}
\caption{Figure 3 is a histogram which shows the distribution of the $\sigma$$_{\circ}$ of the various merger samples compared with the 
gE's, intermediate-mass ellipticals, and spiral bulges.  The bin size is 25 km s$^{-1}$.  The data are plotted as a fraction of the
total number for each sample. The median and mean $\sigma$$_{\circ}$ of each sample are also listed.}
\end{figure}
}
\clearpage
\indent The results in Table 8 and Figure 3 show that the $\sigma$$_{\circ}$ of the normal mergers and shell elliipticals
lie somewhere between intermediate-mass ellipticals and gE's.  The $\sigma$$_{\circ}$ of the LIRG/ULIRG mergers are more akin to 
intermediate-mass ellipticals.  However, the LIRG/ULIRG show larger $\sigma$$_{\circ}$ than those in earlier studies. 
A K-S test between the earlier LIRG/ULIRG studies and the entire LIRG/ULIRG sub-sample in the current study indicates that at the 
0.01 confidence level they do not share the same parent population.  Yet six mergers from those earlier studies are part 
of the merger sample presented here. \\
\indent This begs the question as to what differences exist between the studies which could produce different results.
There are six mergers from the  \markcite{1986ApJ...310..605L}{Lake} \& {Dressler} (1986) study that are also part of the merger sample presented in this paper.
The same stellar absorption line was  used in both studies to derive the central velocity dispersion.  While the 
derived dispersions are not identical, the average difference between the Lake \& Dressler study and this one
is only $\sim$ 6 km s$^{-1}$.  Two major differences between the previous LIRG/ULIRG studies and the current sample are
the inclusion of objects with two distinct nuclei and the stellar absorption lines used to measure $\sigma$$_{\circ}$.
The previous LIRG/ULIRG work cited used only the CO absorption line at either 1.63 or 2.29 $\micron$, whereas this study 
has relied primarily  on the Ca triplet line at 8500 {\AA}.  A comparison of the $\sigma$$_{\circ}$
derived from both the Ca triplet and CO lines for the objects common to both this and earlier studies show that in all but one case, 
the CO line is smaller by anywhere from a few percent to almost 50$\%$.  Figure 3 illustrates the differences in the distribution between
the Ca triplet and CO derived dispersions for the LIRG/ULIRG mergers.  Even for the two mergers in the present sample, NGC 1614 and NGC 2623, 
for which both Ca triplet and CO data were obtained using the same aperture size and the same telescope,  there is a difference in the derived 
$\sigma$$_{\circ}$.  This suggests that the Ca triplet and CO stellar absorption lines may not be probing the same
stellar populations for some or possibly all of the mergers.  These results may explain why earlier LIRG/ULIRG merger
studies concluded that while many of these objects are as luminous as gE's, their kinematic properties are quite
different.  The LIRG/ULIRG mergers in the present study indicate that they are not  consistent with gE's at only the 0.1 confidence level whereas earlier studies have shown this at the 0.001 confidence level.   A more detailed discussion of the differences between 
the Ca triplet and CO $\sigma$$_{\circ}$ is deferred to section 5.6

\subsection{Mergers on the Fundamental Plane as seen in $\kappa$-space}
\indent BBF92 developed a set of orthogonal transformations which allows the Fundamental Plane to be viewed directly in 
terms of mass and mass-to-light within the effective radius:

\begin{equation} {\kappa_{1}} \, {\propto} \, log \, (M/k_{2}) \end{equation}
\begin{equation} {\kappa_{2}} \, {\propto} \, log \, (k_{1}/k_{2}) \, (M/L) \, I_{\rm e}^{3} \end{equation}
\begin{equation} {\kappa_{3}} \, {\propto} \, log \, (k_{1}/k_{2}) \, (M/L) \end{equation}

where k$_{1}$, k$_{2}$, and k$_{3}$ are structural constants and {\it I}$_{\rm e}$ is measured in units of {\it L}$_{\odot}$ pc$^{-2}$ at
{\it K}-band.  The orthogonal transformations are not perfect.  The $\kappa$-space transformation is a 
slightly tilted view of the Fundamental Plane.  Thus, ellipticals on the Fundamental Plane show a slightly 
wider dispersion in their placement in $\kappa$-space. \\
\indent Figure 4 shows the $\kappa$$_{1}$-$\kappa$$_{3}$ space for both the mergers and the P99 ellipticals.
The symbols are the same as Figure 1.  $\kappa$$_{1}$ and $\kappa$$_{3}$ are also given in quantities
of the effective mass (in solar units) and effective {\it M/L} (in solar units).  These conversions
are derived from Appendix A of \markcite{1997AJ....114.1365B}{Burstein} {et~al.} (1997).
The dotted line is the best-fit line from P98.  The mergers show a strong correlation between $\kappa$$_{1}$
and $\kappa$$_{3}$.  The Spearman Rank Correlation Coefficient shows a strong correlation at confidence level of 0.001.
The slope for the P99 ellipticals is 0.147 compared with 0.447 for the mergers.  The mergers appear to have nearly the 
same overall distribution in effective mass as elliptical galaxies, but very different distributions of {\it M/L}.
The presence of a ``tail'' noted earlier in the distribution of the mergers on the Fundamental Plane 
is illustrated in a more physical sense here, and is likely responsible for the steeper slope of the mergers. A K-S test comparing the
({\it M/L})$_{\rm eff}$ ratio of the merger sample as a whole and the ellipticals shows that at a confidence level of 0.001
the two do not arise from the same parent population.  The results are the same when comparing the
normal and LIRG/ULIRG mergers with the P99 ellipticals.  The confidence level is only 0.1 between 
the shell ellipticals and P99 ellipticals.  Ellipticals have a much higher ({\it M/L})$_{\rm eff}$ ratio for the same 
given mass. \\
\clearpage
{
\begin{figure}
\plotone{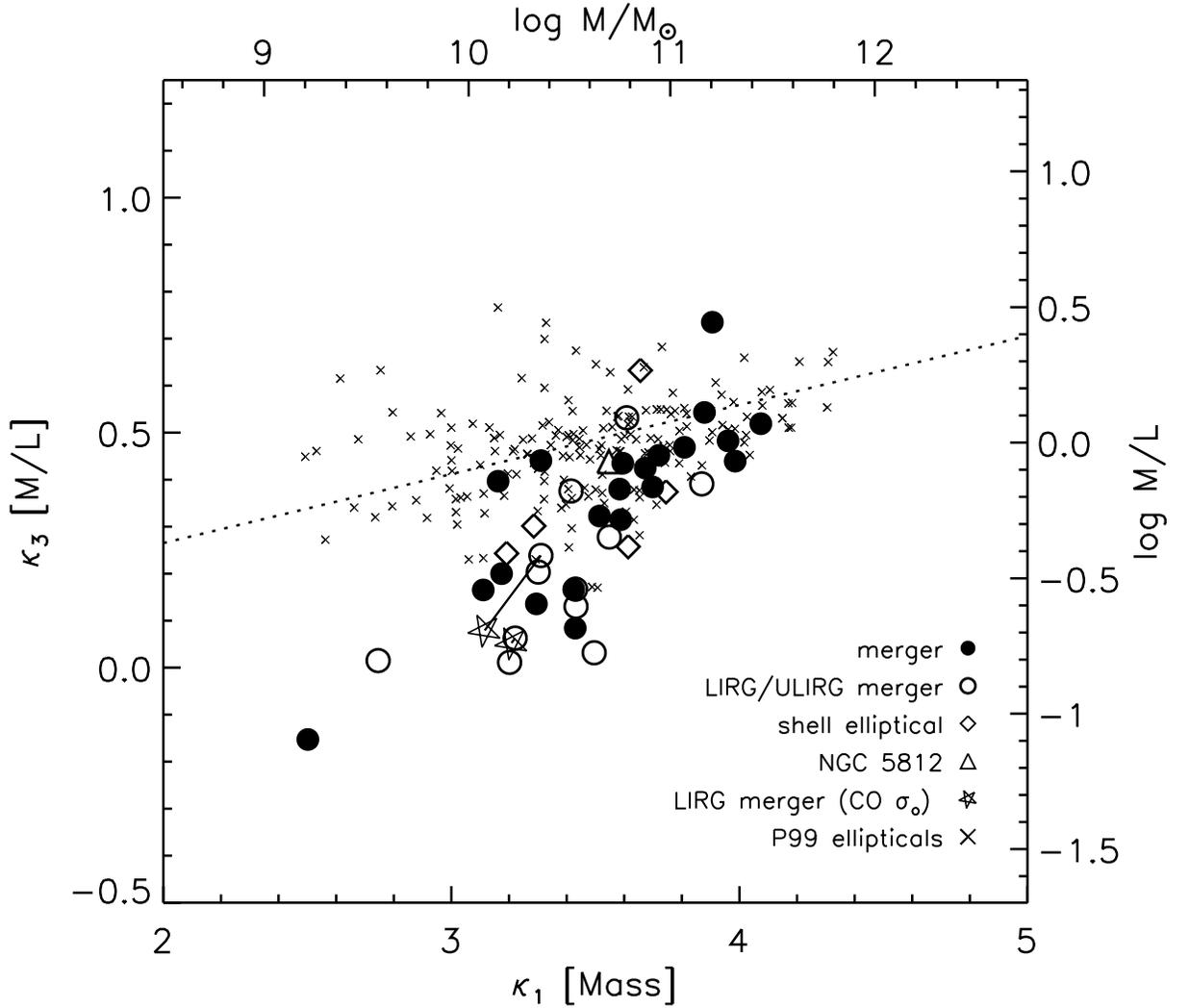}
\caption{$\kappa$-space projection 
of the Fundamental 
Plane as first formulated by BBF92.  The axes are orthogonal transformation of the more familiar 
Fundamental Plane parameters such that $\kappa$$_{1}$ $\propto$ log {\it M}/{\it M}$_{\odot}$ and 
$\kappa$$_{3}$ $\propto$ {\it M/L}$_{\rm eff}$.  The symbols are the same as in Figure 1.  The dotted 
line is the best-fit to the early-type galaxies from P98.  The merger samples shows an obvious 
difference in the {\it M/L}$_{\rm eff}$ ratio as compared with the P99 ellipticals.}
\end{figure}
}
\clearpage
\indent The scatter of some mergers off of the Fundamental Plane has been shown to be the result of photometric 
differences.  This result is borne out in the $\kappa$-space projection as well.  
The differences in the the ({\it M/L})$_{\rm eff}$ ratios at a given mass may be evidence 
of strong star-formation occurring in the mergers. \markcite{1996ApJ...466..114Z}{Zepf} \& {Silk} (1996) modeled the changes 
in the scaling of {\it M/L}  versus {\it M} for elliptical galaxies.  They found that the slope for 
elliptical galaxies may be explained by a burst of star-formation  The ``tail''-like structure
in both the Fundamental Plane and $\kappa$$_{1}$-$\kappa$$_{3}$ diagrams is dominated by
the LIRG/ULIRG galaxies, which are known to contain vast quantities of molecular gas.  These
objects are also known to produce super-starbursts.   \markcite{2003A&A...402..425M}{Mouhcine} \& {Lan{\c c}on} (2003) modeled the photometric evolution of single
stellar populations from 50 Myr to 15 Gyr over a range of metallicities.  One of the goals was to model
the evolution of the stellar mass-to-light ratio in the {\it V}-band and {\it K}-band.  The models
followed the zero age main sequence (ZAMS) to the thermally pulsing regime of the asymptotic giant branch
phase (TP-AGB).  The results showed as the population evolved, the {\it K}-band {\it M}/{\it L} increased.
However, a flattening did occur in the evolution of {\it M}/{\it L} only at {\it K}-band as a result 
of the presence of TP-AGB stars.  Thus, a younger stellar population should produce a smaller {\it M}/{\it L}
ratio for mergers with similar mass to a given elliptical galaxy.  As the central
stellar population ages, the merger may move vertically towards higher values of ({\it M/L})$_{\rm eff}$.  This suggests that
the mergers which lie off of the Fundamental Plane may be undergoing a strong starburst which skews their true position
on the Fundamental Plane.  Kinematically, they are consistent with elliptical galaxies, and as the starburst subsides
then they may move towards the Fundamental Plane as their {\it M/L} ratios increase.

\subsection{The Kormendy Relation}
\indent As noted earlier, differences in the surface brightness and effective radius appear to be responsible
for any offsets between the mergers and the Fundamental Plane (save for NGC 4004, which has
a very small $\sigma$$_{\circ}$).  The rotated view of the Fundamental Plane and orthogonal transformation
to {\it M} and {\it M/L} also appear to confirm this idea.  There also exists a correlation between
{\it R}$_{\rm eff}$ and $<$$\mu$$_{K}$$>$$_{\rm eff}$, first laid out by \markcite{1977ApJ...218..333K}{Kormendy} (1977).
This is essentially a photometric projection of the Fundamental Plane.  Figure 5 is plot of this relation.  
The symbols are the same as in Figure 1.  Unlike the previous figures, data for {\it all} 51 mergers
in the sample are plotted because it does not require any kinematic information.  A Spearman Rank 
Correlation indicates at the 0.001 confidence level that the mergers show a correlation between
log {\it R}$_{\rm eff}$ and $<$$\mu$$_{K}$$>$$_{\rm eff}$.
The best-fit slope of the mergers is 0.126 compared with 0.244 for the P99 ellipticals.  
The range in $<$$\mu$$_{K}$$>$$_{\rm eff}$ is much broader than the P99 ellipticals, extending from 
19.25 mag arcsec$^{-2}$ to 10.68 mag arcsec$^{-2}$.  However, the range in log {\it R}$_{\rm eff}$ 
is much more limited. Unlike $<$$\mu$$_{K}$$>$$_{\rm eff}$ , there is no scatter in {\it R}$_{\rm eff}$ beyond 
that of the P99 ellipticals.  Overall, the mergers seem to show a scatter towards higher $<$$\mu$$_{K}$$>$$_{\rm eff}$
than the P99 ellipticals.  Furthermore, nearly all of the LIRG/ULIRG mergers lie to the right of the relation,
indicating their $<$$\mu$$_{K}$$>$$_{\rm eff}$ are much higher than the other two sub-samples.
\clearpage
{
\begin{figure}
\plotone{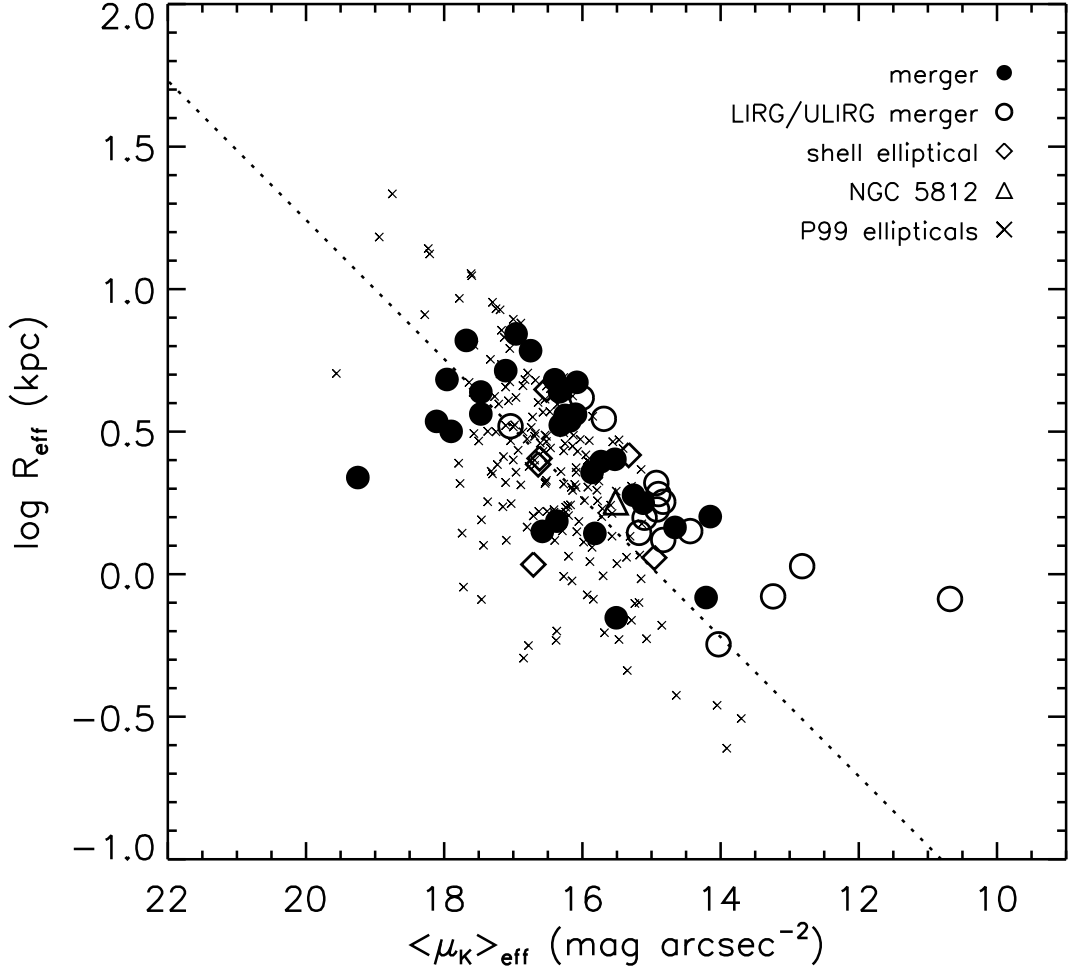}
\caption{The {\it K}-band ``Kormendy Relation'' between 
log {\it R}$_{\rm eff}$ and $<$$\mu$$_{K}$$>$$_{\rm eff}$.  The symbols are the same as Figure 1. 
The dotted line is the best-fit to the early-type galaxies from P98.   A Spearman Rank 
Correlation indicates at the 0.001 confidence level that the mergers show a strong correlation between
log {\it R}$_{\rm eff}$ and $<$$\mu$$_{K}$$>$$_{\rm eff}$.}
\end{figure}
}
\clearpage
\subsection{The Mg$_{2}$-$\sigma$$_{\circ}$ Relation}
\indent The correlation between Mg$_{2}$ and $\sigma$$_{\circ}$ has implications regarding
both the age and metallicity of elliptical galaxies.  It has been known for some time that Mg$_{2}$ shows 
a correlation with $\sigma$$_{\circ}$ \markcite{1981ApJ...246..680T,1981MNRAS.196..381T}({Tonry} \& {Davis} 1981; {Terlevich} {et~al.} 1981).  
Elliptical galaxies with larger velocity dispersions show stronger Mg$_{2}$ strength.
This correlation is important to investigate because the evidence so far indicates that differences in stellar populations 
and age may account for the photometric differences between mergers and elliptical galaxies.
This is the first time that a large sample of mergers has been tested to determine whether they show 
the same correlation between Mg$_{2}$ and $\sigma$$_{\circ}$.  Table 7 lists the 
Mg$_{2}$ magnitudes for each merger.  Figure 6 shows the Mg$_{2}$-$\sigma$ 
correlation.  The dotted line is the best-fit to the Pahre elliptical sample from P98.  The mergers 
do not show the same correlation.  A K-S test shows at the 0.001 confidence level that the P99 sample of ellipticals and mergers
do not arise from the same parent population.  The Spearman Rank Correlation Coefficient indicates that mergers
show a correlation between Mg$_{2}$ and $\sigma$$_{\circ}$ at the 0.01 significance level.
The figure shows nearly a bi-model distribution of the sample between those with negligible
values of Mg$_{2}$ and those approaching similar values to elliptical galaxies.  Moreover,
{\it all} of the LIRG/ULIRG mergers lie far from the best-fit elliptical relation.  \\
\clearpage
{
\begin{figure}
\plotone{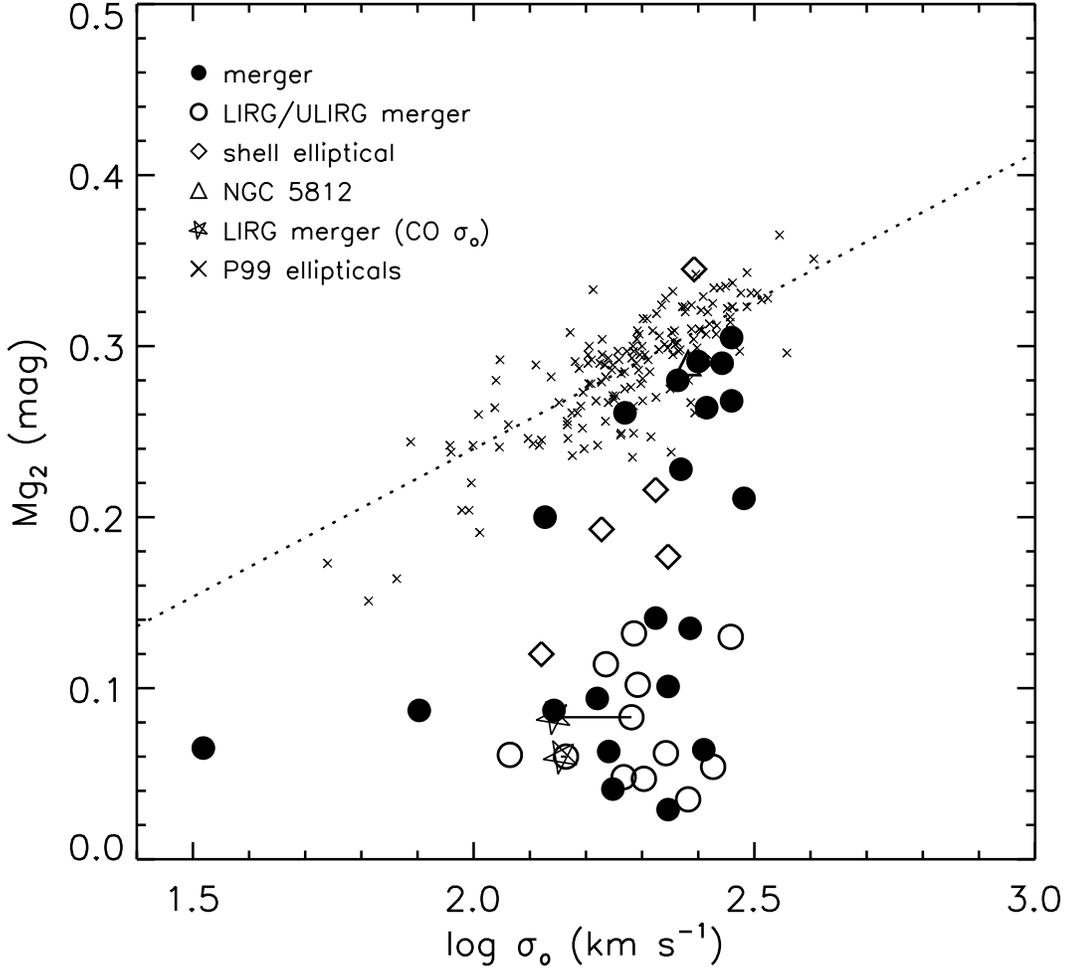}
\caption{The Mg$_{2}$-$\sigma$$_{\circ}$ correlation 
for elliptical galaxies.  The symbols are the same as Figure 1.  The dotted line is the best-fit to the 
early type galaxies in P98.  The mergers show a very large scatter, and no correlation between the
parameters.  The Spearman Rank Correlation Coefficient indicates that mergers
show a correlation between Mg$_{2}$ and $\sigma$$_{\circ}$ at the 0.01 significance level.}
\end{figure}
}
\clearpage
\indent The lack of a strong correlation for mergers between mass and metallicity at first appears to be at odds
with the strong correlations shown for the Fundamental Plane and Kormendy Relation, as well as their kinematic
similarities to elliptical galaxies.  However, the presence of strong star-formation may dilute the Mg$_{2}$ absorption line.
\markcite{1992ApJ...389L..49D}{de Carvalho} \& {Djorgovski} (1992) found that field ellipticals showed a larger scatter than cluster ellipticals in parameter correlations sensitive
to stellar population differences, such as the Mg$_{2}$ index.  \markcite{1998ApJ...508L..43F}{Forbes}, {Ponman}, \&  {Brown} (1998) showed that the scatter in the 
Mg$_{2}$ index is affected by the presence of a young burst population using instantaneous 
burst models from \markcite{1993ApJ...405..538B}{Bruzual} \& {Charlot} (1993).  Strong emission lines, in particular from nearby H$\beta$ and [\ion{O}{3}] lines 
can affect the shape and depth of the Mg$_{2}$ line.  Continuum emission from young stars can also fill in the Mg lines, 
producing a smaller index.  \\
\indent A similar conclusion was reached by \markcite{2004A&A...423..833M}{Michard} \& {Prugniel} (2004) in their study
of peculiar ellipticals.  They found that most of the peculiar ellipticals in their sample had smaller than expected Mg$_{2}$
indices.  These objects were found to have optical colors and spectral line indices consistent with the presence of a younger
stellar population.  However, their results also showed that $\sim$ one-third of the peculiars have normal or even reddish
stellar populations.  They concluded that this is likely the result of either interactions between old galaxies with ISM's 
insufficient to produce new stars or a mixture of medium age and high metallicity stars.  Overall, the more extreme the peculiarity,
the more likely that the population is younger.  This does appears to fit within the paradigm for mergers-make-ellipticals.
As the newly formed ellipticals lose their ``scars'' from the merger, the stellar populations become more consistent with
typical elliptical galaxies and the scatter from the Fundamental Plane decreases.  Figure 7 is a plot of the Mg$_{2}$ index
plotted against the {\it K}-Band Fundamental Plane residuals 
($\Delta$FP = log {\it R}$_{\rm eff}$ - 1.53 log $\sigma$$_{\circ}$ + 0.314 $<$$\mu$$_{K}$$>$$_{\rm eff}$ - 8.30).  Overplotted
are the P99 ellipticals.  A Spearman Rank Correlation Coefficient indicates at the 0.001 confidence level that $\Delta$FP 
is anti-correlated with Mg$_{2}$.  The P99 ellipticals show a much weaker correlation at the 0.05 confidence level.  
\markcite{1996MNRAS.280..167J}{Jorgensen}, {Franx}, \&  {Kjaergaard} (1996) found a similar weak correlation at optical wavelengths, however, they attributed it to a
``left-over'' correlation with $\sigma$$_{\circ}$.  In the case of the mergers, the correlation between Mg$_{2}$ and $\sigma$$_{\circ}$
is significantly weaker than the correlation between $\Delta$FP and Mg$_{2}$.  Thus, it is possible that the correlation in Figure 7
shows that any offset of the mergers from the Fundamental Plane is the result of differences in the stellar population with elliptical galaxies.
A more detailed treatment of the stellar populations of the merger sample using diagnostics such as the Lick Indices and
population synthesis models should be able to confirm or reject this possibility.
\clearpage
{
\begin{figure}
\plotone{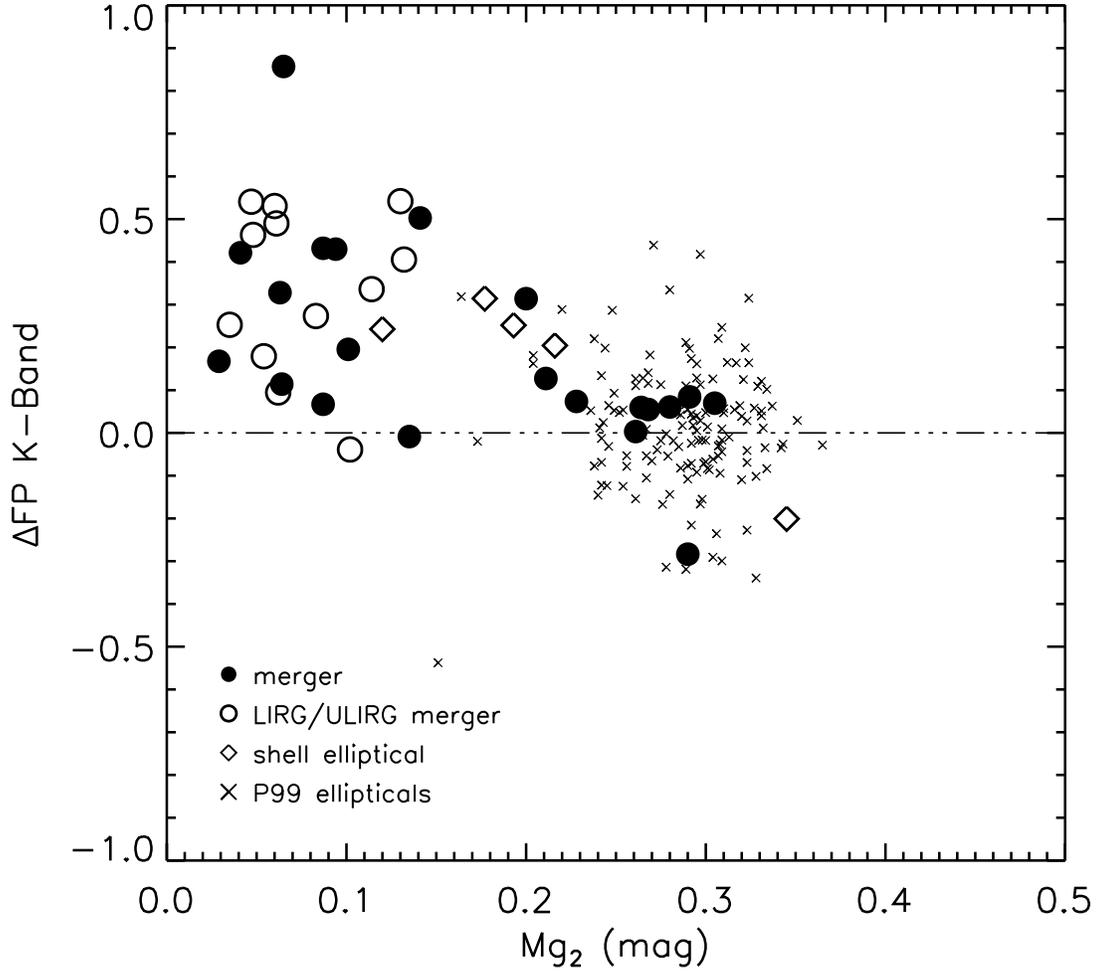}
\caption{Comparison of the {\it K}-Band FP residulals with Mg$_{2}$.
The symbols are the same as Figure 1.  The Spearman Rank Correlation Coefficient indicates that mergers show a strong 
anti-correlation (at the 0.001 confidence) between their distance from the Fundamental Plane and the Mg$_{2}$ index.
The smaller the index, the more offset from the Fundamental Plane.  The P99 ellipticals show a very weak
correlation at the 0.05 confidence level.}
\end{figure}
}
\clearpage
\subsection{The Phase-Space Density of Mergers}
\indent A serious criticism of the merger picture which emerged early on is the discrepancy between the
central densities of disk galaxies and ellipticals.  Merging two stellar disks cannot bring the
central phase-space density to the same levels observed in most elliptical galaxies \markcite{1987nngp.proc.....F}({Gunn} 1987).
\markcite{1986ApJ...310..593C}{Carlberg} (1986) showed that phase-space densities increase as {\it L} decreases
({\it f}$_{c}$ $\propto$ {\it L}$^{-2.35}$).  Using a cooling diagram, \markcite{1989ApJ...342L..63K}{Kormendy} (1989) showed that the amount of 
dissipation needed to produce ellipticals increases as the luminosity decreases.  Only high-luminosity disks
can form the the brightest elliptical galaxies.  \markcite{1992ApJ...390L..53K}{Kormendy} \& {Sanders} (1992) suggested that ULIRG mergers can 
overcome this problem. Observations have shown that for some of these objects the mass in gas is equal to
the mass in stars in the central kpc \markcite{1991ApJ...366L...5S,1991ApJ...366L...1S}({Scoville} {et~al.} 1991; {Sargent} \& {Scoville} 1991) and are
more centrally concentrated than in spiral galaxies \markcite{1991A&A...251....1C}({Casoli} {et~al.} 1991). This would suggest
a strong dissipation of gas to the central region of the merger which can trigger a starburst 
\markcite{1991ApJ...370L..65B,1994ApJ...437L..47M}({Barnes} \& {Hernquist} 1991; {Mihos} \& {Hernquist} 1994).  Even non LIRG/ULIRG mergers are known
to contain large quantities of gas \markcite{1990A&A...228L...5D,1996AJ....111..655H}({Dupraz} {et~al.} 1990; {Hibbard} \& {van Gorkom} 1996).
It has been shown observationally that most mergers undergo strong star-formation 
\markcite{1987nngp.proc...18S,1985MNRAS.214...87J}({Schweizer} 1987; {Joseph} \& {Wright} 1985).
While numerical models predict that a starburst triggered by gaseous dissipation should increase the phase-space
densities of mergers, the observational evidence to support this has been lacking.  \\
\indent Figure 8 shows a plot of the effective phase-space density ({\it f}$_{\rm eff}$) for the mergers
and the P99 sample of elliptical galaxies.  In addition, the bulges of spiral galaxies
are also plotted to compare with the mergers.  The data for the spiral bulges come from 
\markcite{2001A&A...368...16M}{M{\" o}llenhoff} \& {Heidt} (2001) and \markcite{2002MNRAS.335..741F}{Falc{\' o}n-Barroso} {et~al.} (2002).  The data from \markcite{2002MNRAS.335..741F}{Falc{\' o}n-Barroso} {et~al.} (2002)
did not include {\it M}$_{K}$ for the bulges.
{\it M}$_{K}$ for the bulges was derived by first taking {\it M}$_{R}$ from \markcite{2003A&A...405..455F}{Falc{\' o}n-Barroso} {et~al.} (2003)
and then using the ({\it R}-{\it K}\,) colors of each of the bulges 
from \markcite{1997NewA....1..349P}{Peletier} \& {Balcells} (1997) to convert {\it M}$_{R}$ to {\it M}$_{K}$.
The dotted box labeled ``Spirals'' approximates the position of late-type disk galaxies in the 
{\it f}$_{\rm eff}$-{\it M}$_{\rm K}$ plane.  The values for late-type galaxies were taken from 
\markcite{1998MNRAS.296..847M}{Mao} \& {Mo} (1998) and converted from {\it B}-band to {\it K}-band using ({\it B}-{\it K}\,) 
colors from \markcite{2003A&A...410..461G}{Girardi} {et~al.} (2003).
In addition to simple color conversion, the scale length and vertical scale height used by 
Mao \& Mo (both of which 
are factors in approximating an effective radius for late-type galaxies) must be adjusted from {\it B}-band
to {\it K}-band because they are larger at {\it B}-band by a factor of $\sim$ 1.2-2$\times$ 
\markcite{1994A&AS..108..621P}({Peletier} {et~al.} 1994). All data has been corrected to {\it H}$_{\circ}$ = 75 km s$^{-1}$ Mpc$^{-1}$.\\
\indent  The parameter {\it f}$_{\rm eff}$ was defined by \markcite{1993ApJ...416..415H}{Hernquist} {et~al.} (1993)
as way of estimating the phase-space density from observed parameters:
\begin{equation} f_{eff} \, \equiv \, \left[\frac{1}{\sigma_{\circ} \, R^2_{\circ}}\right] \end{equation}
It is similar to the phase-space density used by \markcite{1986ApJ...310..593C}{Carlberg} (1986) except it replaces 
{\it r}$_{\rm c}$ from a King profile with {\it r}$_{\rm eff}$, the half-light radius derived from a 
de Vaucouleurs {\it r}$^{1/4}$ fit to the surface brightness profile.  The results in Figure 8 show that
phase-space densities of the mergers are very close to those of elliptical galaxies.  This suggests 
that the mergers have undergone dissipative collapse.
\clearpage
{
\begin{figure}
\plotone{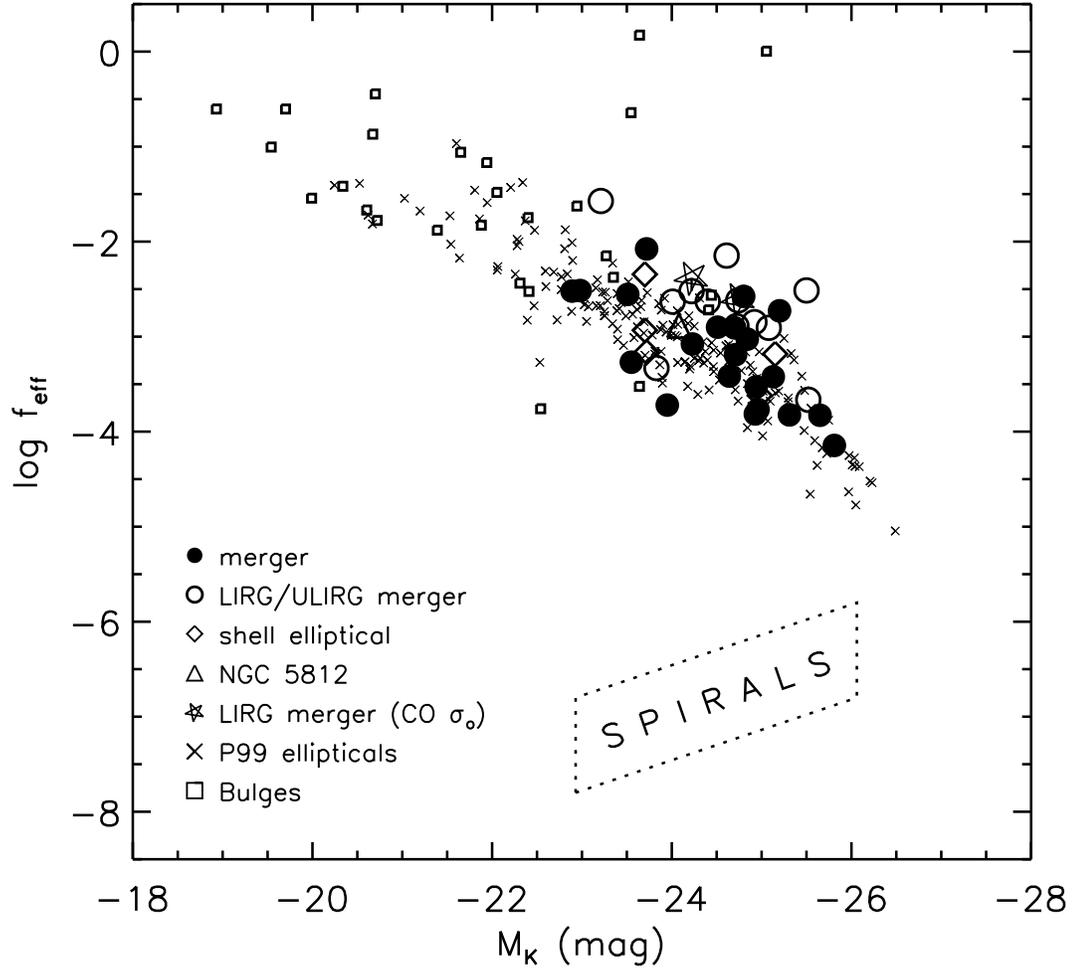}
\caption{Plot of the effective phase-space (f$_{\rm eff}$) defined by Hernquist et al. (1993).
The symbols are the same as Figure 1, however, they also include a sample of bulges (squares) as a 
comparison.  In general, the f$_{\rm eff}$ of the merger sample is consistent with elliptical galaxies.
The LIRG/ULIRG mergers do have a higher f$_{\rm eff}$ than the other mergers in the sample.}
\end{figure}
}
\clearpage
\indent A K-S test comparing the {\it f}$_{\rm eff}$ of the mergers with the P99 ellipticals
indicates that two populations likely share the same parent population.   The same results
are found when the normal merger and shell elliptical sub-samples are compared with the P99
ellipticals.  However, the null hypothesis is rejected at the 0.1 confidence level between the LIRG/ULIRG
mergers and P99 ellipticals.  The LIRG/ULIRGs in Figure 8 have a higher phase-space density compared with
the normal mergers, shell ellipticals, and P99 ellipticals. 

\subsection{Comparison of Ca triplet and CO velocity dispersions}
\indent As noted in Section 5.1, there are significant differences between the velocity dispersions 
derived from the Ca triplet at $\sim$ 8500 {\AA} and the CO line at 2.29 $\micron$.  These differences
have led to different conclusions on the end-state of mergers, in particular, LIRG/ULIRG mergers.
Even within this study here, one of the two mergers with both Ca triplet and CO measurements shows
significant differences in the derived $\sigma$$_{\circ}$. Discrepancies in data reduction and analysis between 
the Ca triplet and CO absorptions lines can be ruled out.
The galaxies were measured using the same 1.53 kpc aperture size in the spatial direction, as well 
as with identical position angles.  The slit width
used for the Ca triplet line was 0{$\arcsec$}.5 and was 0{$\arcsec$}.432 for the CO line. Both the Ca triplet 
and CO lines were fit in exactly the same manner, using a Gauss-Hermite polynomial fit in pixel-space to 
determine the LOSVD.  The continuum were normalized in a similar fashion.  The only difference in the 
fitting routing were the template stars used.  The best fit template star using the Ca triplet line for NGC 1614 was a G0III
star, HD 332389 and for NGC 2623 a G8III star, HD 100347.  In the near-infrared, the best fit template star
for NGC 1614 was an M7.5III, Rx Boo and for NGC 2623 an M0III star, $\lambda$ Dra.  M-giant template stars
were tested with the Ca triplet lines for NGC 1614 and NGC 2623 and found to be 2-3 times poorer fits than the
G-type giants. The earliest CO template star observed was a K0III.  However, the depth of the CO line in the K0III star 
was found to be too shallow as compared with the depth of the CO line for both NGC 1614 and NGC 2623.\\
\indent Table 9 lists mergers in the present sample for which CO derived $\sigma$$_{\circ}$ also exist in the 
literature.  Only 1 merger, Arp 193,  has a larger CO derived $\sigma$$_{\circ}$.  A K-S test between 
the Ca triplet and CO derived dispersions rejects at the 0.025 
confidence level that the two populations arise from the same parent population, even though the mergers are identical.\\
\clearpage
{
\begin{deluxetable}{lcc}
\tabletypesize{\normalsize}
\setlength{\tabcolsep}{0.1in}
\tablewidth{0pt}
\tablenum{9}
\pagestyle{empty}
\tablecolumns{3}
\tablecaption{Comparison of Ca triplet and CO derived $\sigma$$_{\circ}$}
\tablehead{
\colhead{Merger} &
\colhead{Ca Triplet $\sigma$$_{\circ}$} &
\colhead{CO $\sigma$$_{\circ}$} \\
\colhead{Name} &
\colhead{km s$^{-1}$}&
\colhead{km s$^{-1}$}\\
}
\startdata
NGC 1614      &146  &143,\tablenotemark{1},75\tablenotemark{2},154\tablenotemark{4}\\
NGC 2623      &191  &139,\tablenotemark{1}95\tablenotemark{3}\\
NGC 3256      &241  &113\tablenotemark{4}\\
NGC 4194      &116  &104\tablenotemark{5}\\
Arp 193       &172  &206\tablenotemark{5}\\
AM 2055-425   &185  &140\tablenotemark{6}\\
NGC 7252      &166,177\tablenotemark{7} &123\tablenotemark{5}\\
IC 5298       &193  &151\tablenotemark{3}\\
\enddata
\tablecomments{(1) this study, (2) \cite{1994ApJ...433L...9S},
(3) \cite{1996ApJ...470..222S}, (4) \cite{1995A&A...301...55O},
(5) \cite{1999MNRAS.309..585J}, (6) \cite{2001ApJ...563..527G},
(7) \cite{1986ApJ...310..605L}}
\end{deluxetable}

}
\clearpage
{
\begin{figure}
\plotone{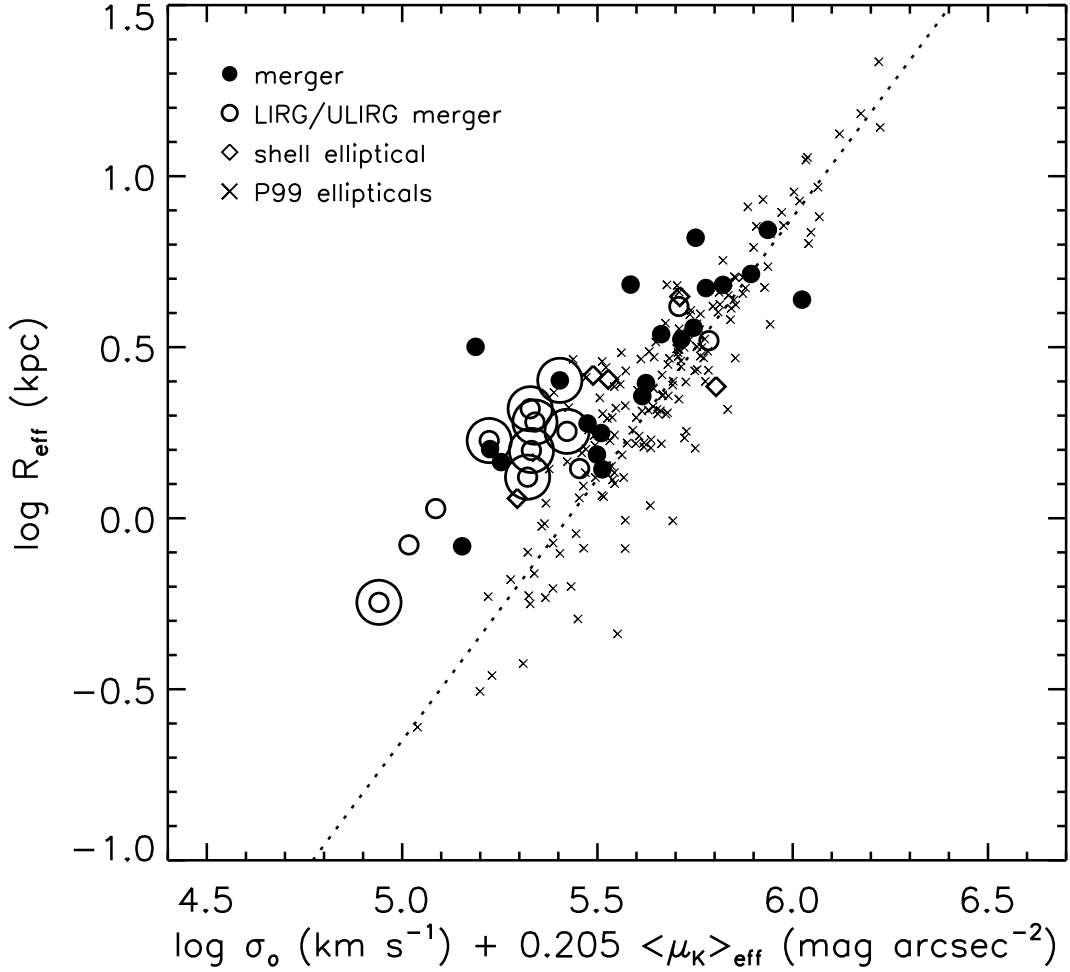}
\caption{The {\it K}-band Fundamental Plane of Elliptical Galaxies as plotted in Figure 1
(symbols have the same meaning).
The mergers in Table 9 are all circled to show where they lie on the Fundamental Plane.
{\it ALL} of these mergers lie in the ``tail'' off of the FP.}
\end{figure}
} 
\clearpage
\indent These results raise questions about earlier merger studies which have relied solely on 
the CO absorption line to derive velocity dispersions.  Are the stellar populations 
observed in the optical and infrared the same?  However, before addressing that, it should be noted 
that {\it all} of the mergers in Table 9 lie in one specific region of the Fundamental Plane in Figure 1.
Figure 9 shows the Fundamental Plane, plotted as it was in Figure 1, with the mergers that 
have both Ca triplet and CO dispersions circled.  The plot shows the mergers with only the Ca triplet dispersions.
The CO dispersions would shift the circled mergers further left (except for Arp 193).
The circled mergers all lie within the ``tail'' noted earlier.  If the CO velocity dispersions 
are used to plot the points, then all but one would shift further to the left.\\
\indent If the mergers listed in Table 9 were distributed uniformly among the merger sample,
then any conclusions drawn about the differences between the Ca triplet and CO absorption lines
could be applied universally to all mergers, and, perhaps, even to some elliptical galaxies.
However, the mergers in Table 9 may not be representative of all mergers.  First, all but one
are LIRG/ULIRGs.  It has already been established that these mergers are somewhat different
from ``normal'' mergers, in particular, they have a higher central surface brightness.\\
\indent \markcite{2003AJ....125.2809S}{Silge} \& {Gebhardt} (2003) also found similar discrepancies for a sample of 25 early-type galaxies.  
They tested for systematic errors from continuum fitting and choice of template stars and found that at most
they could account for $\sim$ 5-10$\%$ decrease in derived dispersions.  Turning to a physical explanation,
they compared Mg$_{2}$ indices with CO equivalent widths and found no correlation.  Furthermore, 
the equivalent widths of the CO lines were higher than expected from K-type giants and more in line with
cool M-type giants. This suggests that the CO absorption line is not probing the same stellar population
as the optical absorption lines.  \markcite{1999A&A...350....9O}{Oliva} {et~al.} (1999) point out that the CO 2.29 $\micron$ band-head is primarily sensitive
to micro-turbulent photospheric motions which increase at lower stellar temperatures and are larger
in red supergiants (RSGs) than giants of a given temperature.  In fact, they further point out that the CO line
is essentially a degenerate diagnostic.  Young stars of low metallicity and old, highly metallic cool red
giants can have similar equivalent widths.  Oliva et al. note difficulties in using
equivalent widths to differentiate between stellar types.  They {\it suggest}
that an EW $\geq$ 15 {\AA} may be indicative of younger stars since that is larger than the EW of 
most metal-rich globular clusters.  However, it is not uncommon to find some starbursts with a 
CO EW $\leq$ 15 {\AA}.  The CO EWs of NGC 1614 and NGC 2623 are 14.4 {\AA} and 12.7 {\AA} respectively
(using the EW definition from \markcite{1993A&A...280..536O}{Origlia}, {Moorwood}, \&  {Oliva} (1993)), which is
close to the borderline, and consistent with either M-type giants or RSGs.  \markcite{1995A&A...301...55O}{Oliva} {et~al.} (1995)
measured both the $\sigma$$_{\circ}$ and the EW of NGC 3256 and derived a value of 12.9 {\AA}.
The Ca triplet EWs (using the CaT* definition from \markcite{2001MNRAS.326..959C}{Cenarro} {et~al.} (2001) ) are 5.1 {\AA}, 6.3 {\AA}, and  6.8 {\AA} for
NGC 1614, NGC 2623, and NGC 3256 respectively, which are more consistent with K or earlier giant stars.\\
\indent This difference in which stellar population is probed depending on the optical or infrared absorption line used may 
be supported by the {\it K}-band photometry.  As noted above, the mergers with both Ca triplet and CO observations lie 
 are primarily all LIRG/ULIRGs which have much higher surface brightnesses compared with 
the other mergers in the sample.  \markcite{2002A&A...393..149M}{Mouhcine} \& {Lan{\c c}on} (2002) modeled the contribution of asymptotic giant branch (AGB) stars 
to the total {\it K}-band luminosity for a single-burst population.  They found that the contribution evolves
rapidly from a few percent at $\simeq$ 0.1 Gyr to $\simeq$ 60$\%$ of the light at 0.6-0.7 Gyr
and then quickly declines soon after to $\leq$ 30$\%$.  The {\it K}-band light from these mergers
may be dominated by AGB stars, which is what the CO absorption line is also detecting.  The other mergers, which
lie closer to, or on the plane, may be at a point where the light from AGB stars does not dominate the {\it K}-band continuum.
It is unlikely that {\it all} of the mergers in the sample lie within the same narrow age range (0.1 Gyr)
in which their light is dominated by contributions from AGB stars.  Thus, it is possible that the CO $\sigma$$_{\circ}$ may eventually
match the Ca triplet line. A more detailed analysis of optical and infrared spectra of the nuclei should be able to 
determine whether these wavelength ranges are detecting the same stellar populations and constrain which
mergers show srong contributions from AGB stars.
Since ESI has a large optical wavelength range, the data used to derive velocity dispersions can also be used
to get a handle on the stellar populations in the optical.  Additional medium resolution ($\lambda$ $\sim$ 1200) 
infrared spectra from 1-2.5$\micron$ have been obtained for $\sim$ two-thirds of the sample using SPEX on the 3.0m IRTF 
and UIST on the 3.8m UKIRT. The optical and IR spectra have approximately the same slit width.  These results will be presented in a subsequent paper.  Additional high-resolution CO observations to derive $\sigma$$_{\circ}$ for mergers which lie on the 
Fundamental Plane would be vital to determining whether the Ca triplet and CO derived dispersions will eventually match
each other. \\
\indent A second trend noted by \markcite{2003AJ....125.2809S}{Silge} \& {Gebhardt} (2003) was a decrease in the fractional differences in velocity 
dispersions when moving from S0 to E galaxies.  This might be the result of dust which is both spatially coincident and 
geometrically aligned with a central disk.  The dust would block the disk at optical wavelengths, yet be transparent in the infrared.  
Thus CO measurements would probe the motions of stars in the disk, producing smaller velocity dispersions.  The presence
of a central stellar disk in mergers is inferred from numerical models which predict the formation of a central gaseous disk
as a result of gaseous dissipation \markcite{1991ApJ...370L..65B,2002MNRAS.333..481B}({Barnes} \& {Hernquist} 1991; {Barnes} 2002).  The {\it K}-band photometry
from Paper I show a significant number of mergers have disky isophotes.  This suggests that the Ca triplet and CO
derived $\sigma$$_{\circ}$ may never match each other. If this is the case, then it is important to consider which 
absorption line is a better measure of the dynamical mass of the system.  If the CO absorption line is dominated
by stars which lie in a central disk which produces smaller $\sigma$$_{\circ}$, then this disk may be de-coupled
from the dynamics of the larger galactic system.  Numerical simulations show that a central gaseous disk can form
which could produce stars that are kinematically decoupled from the larger system.  In other words, the central disk
of stars would have been formed {\it in situ} and have no ``knowledge" of the dynamics of the system as a whole.  In contrast,
the stars contributed from the progenitor spiral galaxies, would be sensitive to the dynamical state of the whole system. 

\section{Summary and Discussion}
\indent Below are a summary of the main results which have emerged from the data presented here:
\indent 1) Most of the mergers lie on or close to the Fundamental Plane (within the same scatter as the P99 ellipticals).
Those mergers which lie beyond the scatter of the P99 ellipticals off of the Fundamental Plane all fall within the same 
region and appear to form a discrete structure.\\
\indent 2) The $\sigma$$_{\circ}$ of the mergers, with the exception of NGC 4004, are entirely consistent with those 
of elliptical galaxies.  Furthermore, the dispersions appear to lie somewhere between values of intermediate-mass
and giant elliptical galaxies.  The $\sigma$$_{\circ}$ of the LIRG/ULIRG mergers are larger than previous observations,
placing them closer to giant ellipticals.\\
\indent 3) The differences between mergers which lie off of the Fundamental Plane and ellipticals appear to be solely 
photometric in nature.\\
\indent 4) Mergers show a very weak mass-metallicity correlation that is quite different from elliptical galaxies.
Furthermore, the Mg$_{2}$ index shows an anti-correlation with the residuals of the FP.  The larger the
$\Delta$FP the smaller the Mg$_{2}$ index.  This suggests that the differences between mergers and ellipticals
may be due to differences in the stellar populations.\\
\indent 5) The central phase space density of mergers appear to be consistent with those of elliptical galaxies.
This suggests that the mergers have undergone dissipation.\\
\indent 6) A significant difference has been found in the derived values of $\sigma$$_{\circ}$ between the Ca triplet
and CO absorption lines.  The differences may be the result of the two absorption lines probing different stellar populations.
A second possibility is that the CO line is probing a central dust enshrouded disk.  The {\it only} way to confidently determine why
a discrepancy exists is to obtain high-resolution CO observations of the mergers which lie on the Fundamental Plane.\\
\indent The overall picture painted by these results suggests that spiral-spiral mergers produce objects which show nearly
the same strong correlations between kinematic and photometric properties as elliptical galaxies (i.e. The Fundamental
Plane).  Any discrepancies between mergers and ellipticals have been shown not to be a function of $\sigma$$_{\circ}$.
This is critical because the characteristic $\sigma$$_{\circ}$ is likely to settle quickly to its final value after
the two nuclei merge.  This parameter will not evolve, or change over time.  The major differences
between mergers and ellipticals could be a function of the stellar populations. The evidence for this are the discrepancies
in surface brightness, {\it M/L}, and metallicity. These suggest that a younger stellar population is superposed in the centers
of the mergers, which increases their central luminosity and contaminates (i.e. ``fills in'') or dominates the observed
metallicities.  The strong anti-correlation between $\Delta$FP and Mg$_{2}$ strengthens this assertion.
A possible secondary piece of evidence to support this is the discrepancies between the Ca triplet
and CO derived $\sigma$$_{\circ}$.  The CO absorption line is sensitive to both young, metal-poor RSGs as well
as older late-type giants.  If a strong starburst is present in the central regions of the mergers, the light from this
could very well dominate the CO lines.  These stars would be newly formed and would have no ``memory'' 
of the kinematics of the original late-type giants contributed from the two progenitors.  This could explain why the $\sigma$$_{\circ}$
are lower than those derived from the Ca triplet, which are probing the late-type giants.  Furthermore, the 
phase-space densities of the mergers are consistent with those of elliptical galaxies.  This supports the notion that these
objects have undergone dissipative collapse, increasing the central stellar densities via a starburst o levels equivalent to 
elliptical galaxies.  Cold or dissipationless merging can only form a giant elliptical galaxy from two massive spiral progenitors.
If the mergers presented here were all the result of dissipationless merging, than their $\sigma$$_{\circ}$  (both Ca triplet
and CO) should {\it all} be consistent with the most massive luminous elliptical galaxies ($\sigma$$_{\circ}$ $\gg$ 250
km s$^{-1}$). \\
\indent The above conclusion is without a doubt painted in rather broad strokes and requires further analysis
and observations to test its validity.  First and foremost is to understand the nature of the Ca triplet and CO discrepancies.
Unfortunately, all of the mergers which have CO measurements lie in an area offset from the Fundamental Plane.
In order to constrain whether the discrepancy is the result of RSGs dominating the light, or a central stellar disk enshrouded
in dust, CO observations are needed of mergers which lie {\it on} the Fundamental Plane.  A more detailed analysis
of the stellar populations using both optical and infrared absorption lines may help explain the lack of a strong
mass-metallicity relationship as well as confirm that the offset of some mergers from the Fundamental Plane
is the result of differences in {\it M/L} produced by younger stellar populations.  

\acknowledgments
We would like to thank John Hibbard for helpful and informative discussions on earlier drafts of the paper, as well as
Brad Whitmore and Roeland van der Marel for subsequent discussions which have improved the structure and content of the paper.
 A special thanks is given to Michael Cushing for his help 
and guidance on developing the IDL code used to measure the velocity dispersions from the extracted spectra.
We would also like to thank Michael Connelley for taking {\it K}-band data 
of one of the mergers.  This research has made use of the NASA/IPAC Extragalactic Database
(NED) which is operated by the Jet Propulsion Laboratory, California Institute of Technology,
under contract with the National Aeronautics and Space Administration.  This research has been supported
in part, by a Fellowship from the NASA Graduate Student Researchers Program, Grant $\#$ NGT5-50396.

\appendix 
\section{Ca triplet and CO spectra}
\indent In this section we present the rest-frame wavelength spectra centered on the Calcium triplet absorption line for 38
mergers and 1 E0 galaxy in the sample, and the CO absorption line for 2 mergers in the sample.  The solid line plotted in each
figure is the galaxy spectrum, the superposed dashed-line is the convolved stellar template.

{
\begin{figure}
\plotone{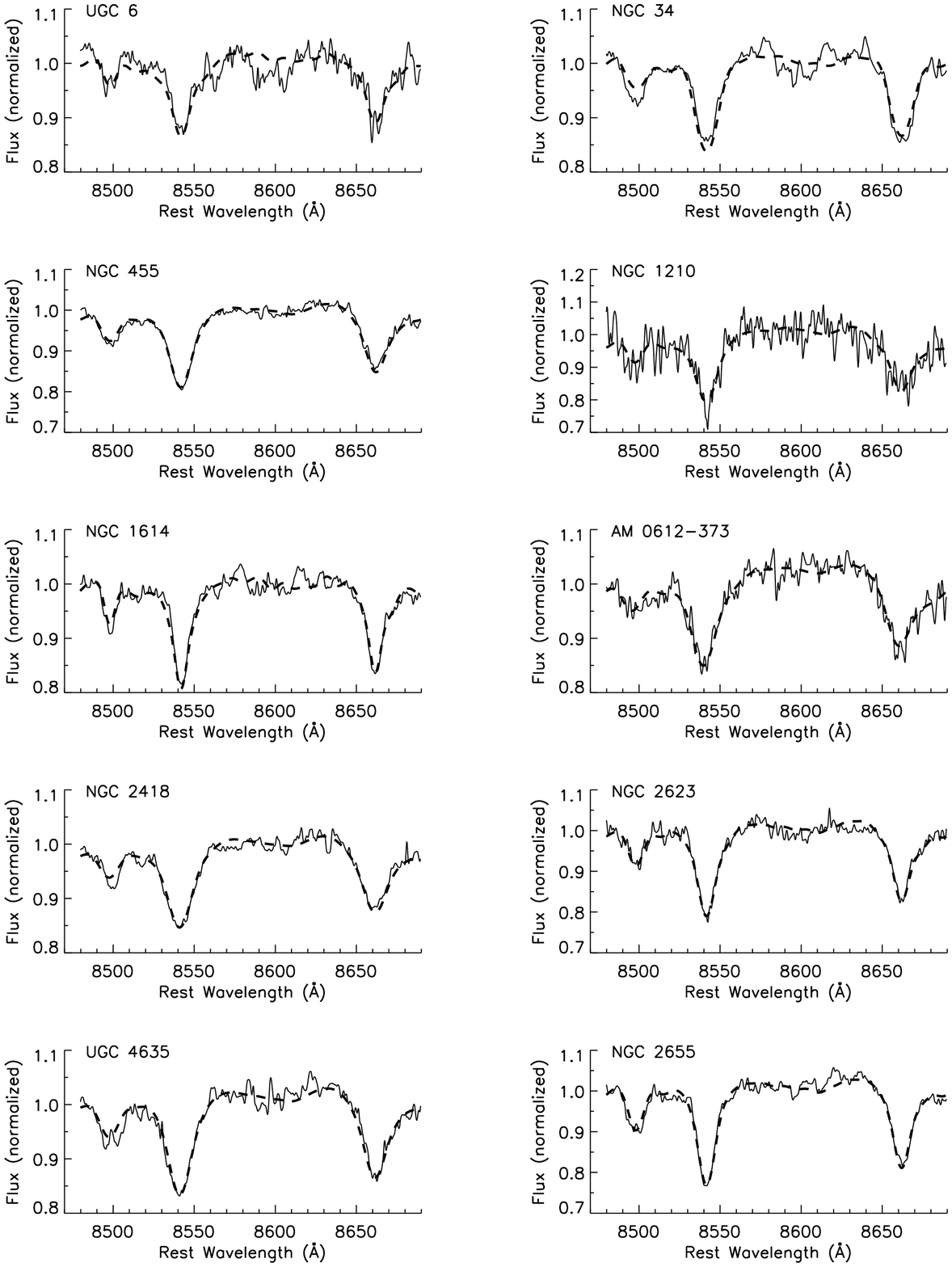}
\end{figure}
} 
{
\begin{figure}
\plotone{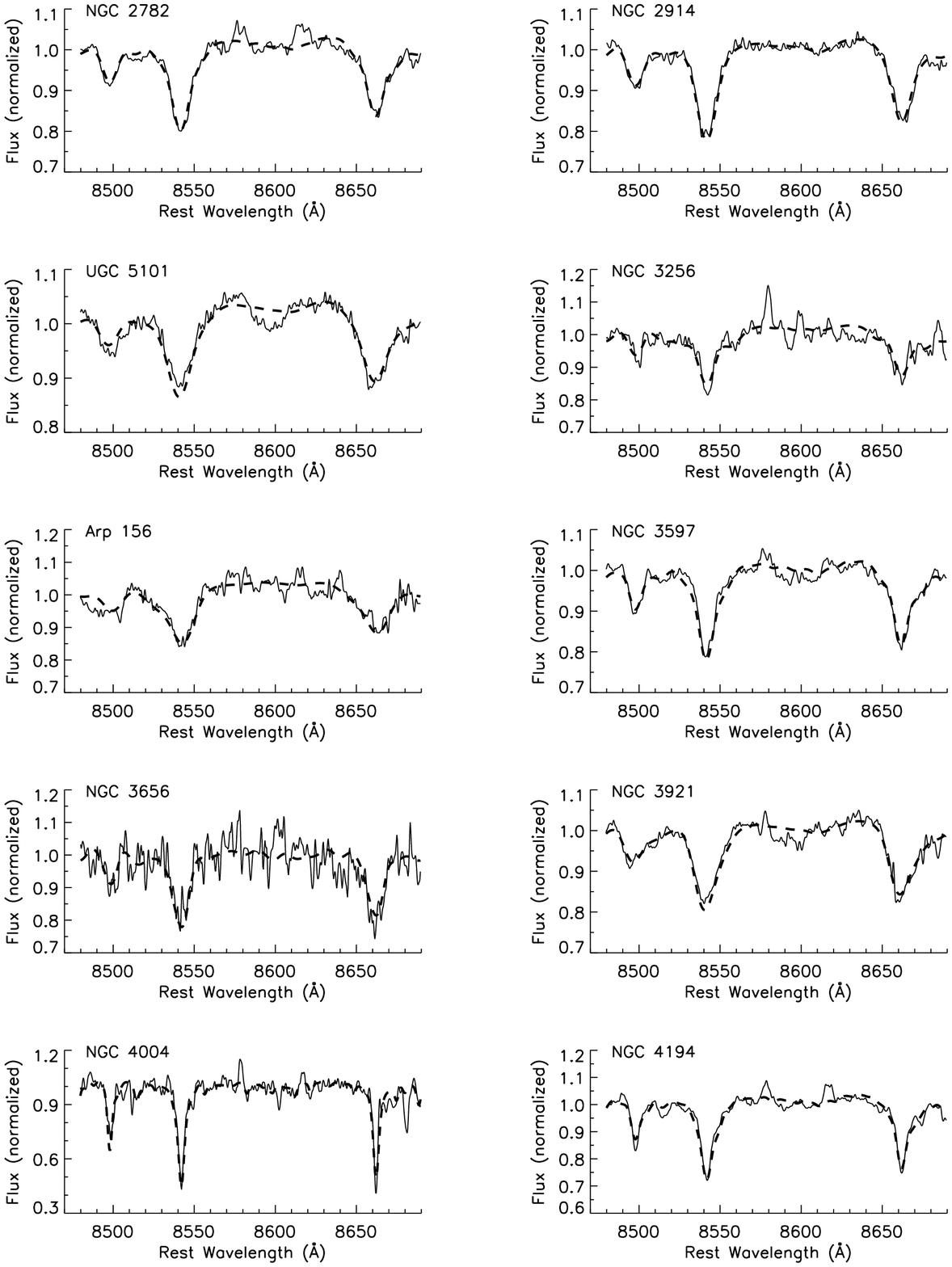}
\end{figure}
} 
{
\begin{figure}
\plotone{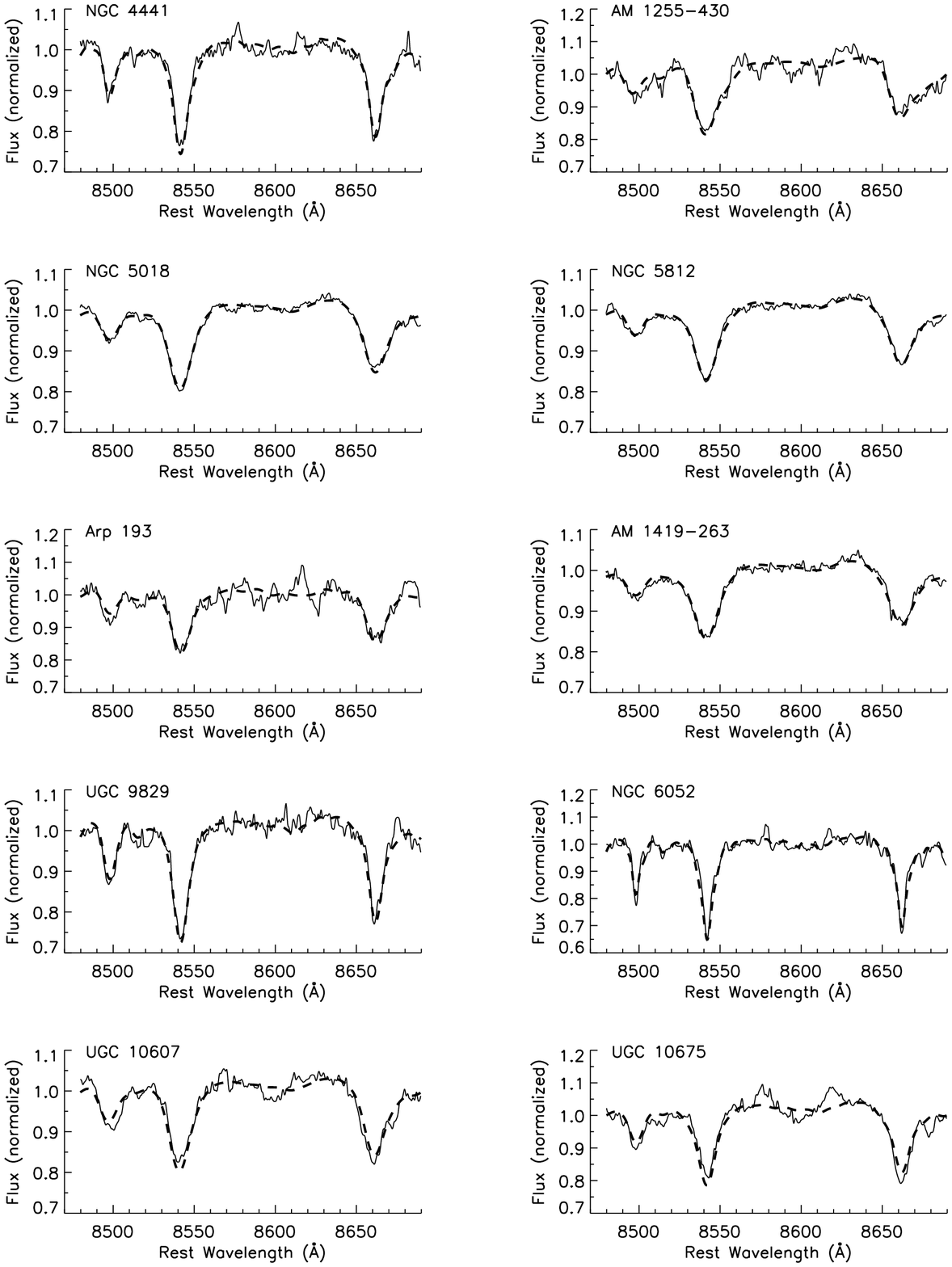}
\end{figure}
} 
{
\begin{figure}
\plotone{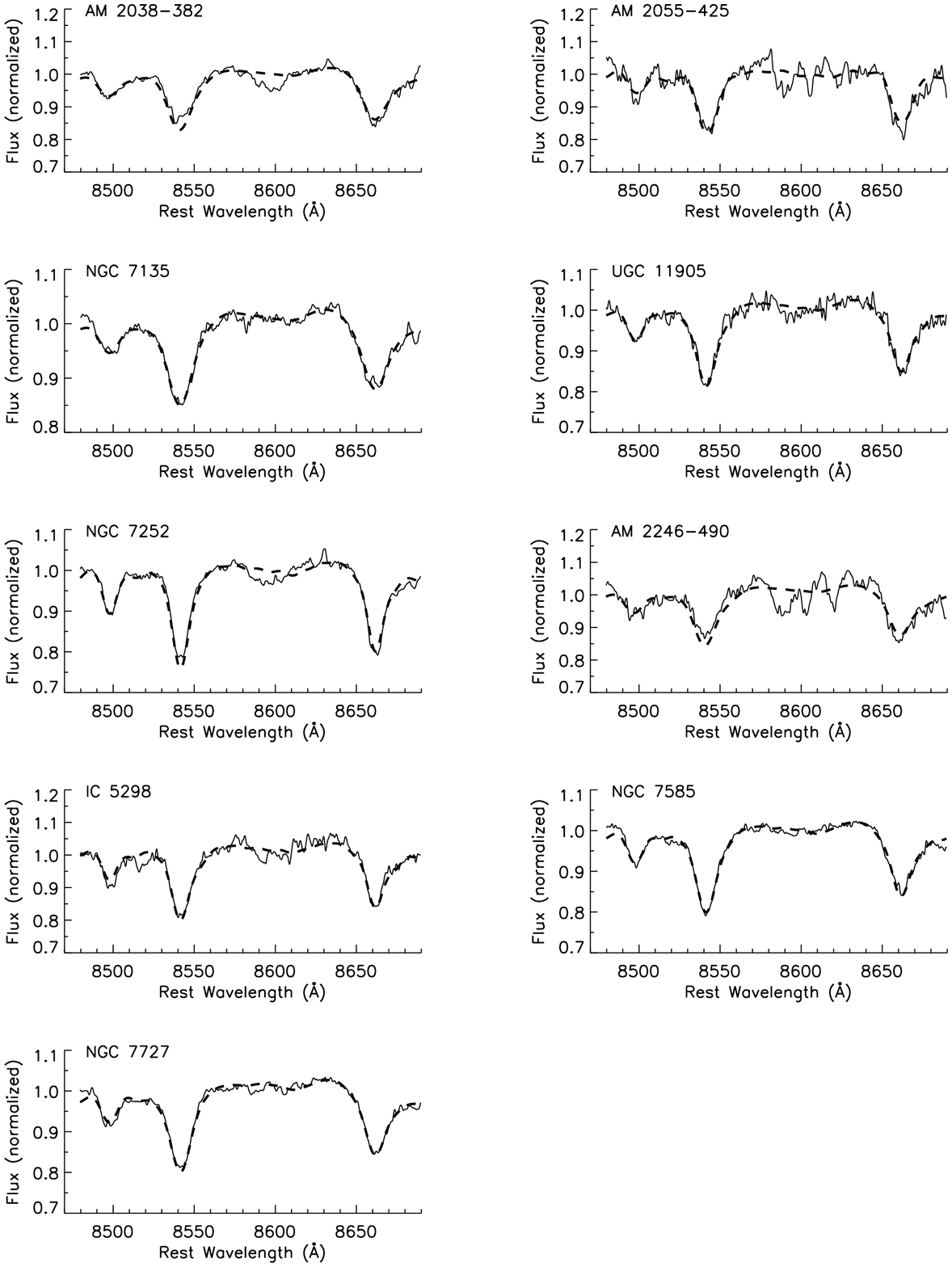}
\end{figure}
} 
{
\begin{figure}
\plotone{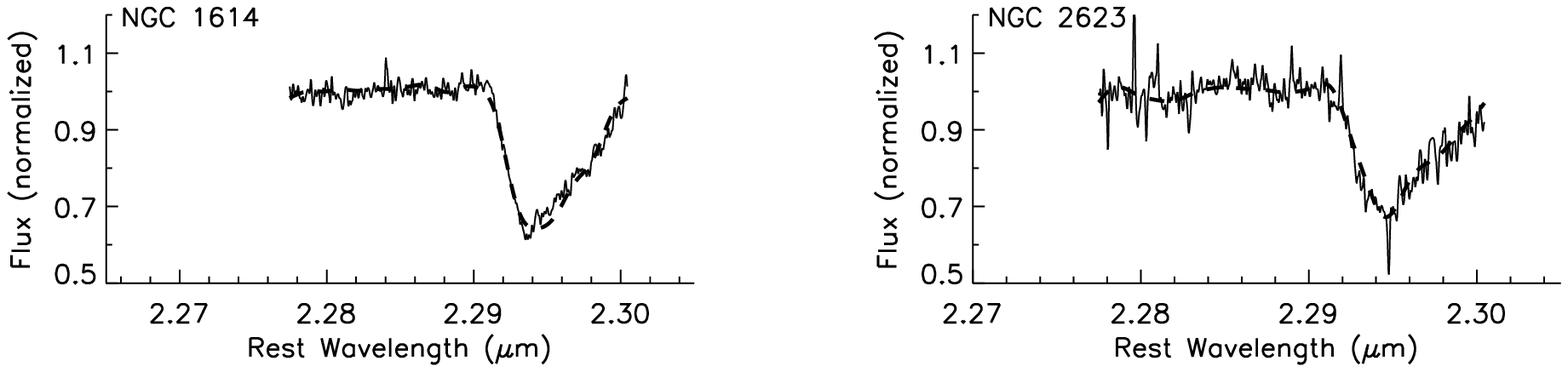}
\end{figure}
} 

\clearpage
\bibliography{}

\end{document}